\documentclass[sigconf, nonacm]{acmart}
\usepackage[utf8]{inputenc}
\usepackage{graphicx}
\usepackage{subfig}
\usepackage{dutchcal}
\usepackage[mathscr]{eucal}
\usepackage{latexsym}
\usepackage{setspace}
\usepackage{commath}
\usepackage[english]{babel}
\usepackage[noend]{algpseudocode}
\usepackage{algorithmicx}
\usepackage{booktabs}
\usepackage{algorithm}
\usepackage[utf8]{inputenc}
\usepackage{mathtools}
\usepackage{centernot}
\usepackage{makecell}
\usepackage{tabularx}
\usepackage{xcolor}
\usepackage{mathtools}
\usepackage{soul}
\usepackage{svg}
\usepackage{hyperref}
\usepackage{comment}

\newcommand\vldbdoi{XX.XX/XXX.XX}
\newcommand\vldbpages{XXX-XXX}
\newcommand\vldbvolume{14}
\newcommand\vldbissue{1}
\newcommand\vldbyear{2020}
\newcommand\vldbauthors{\authors}
\newcommand\vldbtitle{\shorttitle} 
\newcommand\vldbavailabilityurl{https://github.com/szeighami/SNH}
\newcommand\vldbpagestyle{plain}

\newcommand{\sz}[1]{\textcolor{red}{\bf\small [#1 --Sep]}}

\newcommand{\ra}[1]{\textcolor{teal}{\bf\small [#1 --Ritesh]}}
\newtheorem{problem}{Problem}
\algrenewcommand\algorithmicrequire{\textbf{Input:}}
\algrenewcommand\algorithmicensure{\textbf{Output:}}
\begin{document}
\title{A Neural Database for Differentially Private\\ Spatial Range Queries
}

\author{Sepanta Zeighami}
\affiliation{%
  \institution{USC}
}
\email{zeighami@usc.edu}

\author{Ritesh Ahuja}
\affiliation{%
  \institution{USC}
}
\email{riteshah@usc.edu}

\author{Gabriel Ghinita}
\affiliation{%
  \institution{UMass Boston}
}
\email{gabriel.ghinita@umb.edu}
\author{Cyrus Shahabi}
\affiliation{%
  \institution{USC}
}
\email{shahabi@usc.edu}

\begin{abstract}
Mobile apps and location-based services generate large amounts of location data that can benefit research on traffic optimization, context-aware notifications and public health (e.g., spread of contagious diseases). To preserve individual privacy, one must first sanitize location data, which is commonly done using the powerful differential privacy (DP) concept. However, existing solutions fall short of properly capturing density patterns and correlations that are intrinsic to spatial data, and as a result yield poor accuracy. We propose a machine-learning based approach for answering statistical queries on location data with DP guarantees. We focus on countering the main source of error that plagues existing approaches (namely, uniformity error), and we design a neural database system that models spatial datasets such that important density and correlation features present in the data are preserved, even when DP-compliant noise is added. We employ a set of neural networks that learn from diverse regions of the dataset and at varying granularities, leading to superior accuracy. We also devise a framework for effective system parameter tuning on top of {\em public} data, which helps practitioners set important system parameters without having to expend scarce privacy budget. Extensive experimental results on real datasets with heterogeneous characteristics show that our proposed approach significantly outperforms the state of the art.
\end{abstract}
\maketitle


\pagestyle{\vldbpagestyle}
\begingroup\small\noindent\raggedright\textbf{PVLDB Reference Format:}\\
\vldbauthors. \vldbtitle. PVLDB, \vldbvolume(\vldbissue): \vldbpages, \vldbyear.\\
\href{https://doi.org/\vldbdoi}{doi:\vldbdoi}
\endgroup
\begingroup
\renewcommand\thefootnote{}\footnote{\noindent
This work is licensed under the Creative Commons BY-NC-ND 4.0 International License. Visit \url{https://creativecommons.org/licenses/by-nc-nd/4.0/} to view a copy of this license. For any use beyond those covered by this license, obtain permission by emailing \href{mailto:info@vldb.org}{info@vldb.org}. Copyright is held by the owner/author(s). Publication rights licensed to the VLDB Endowment. \\
\raggedright Proceedings of the VLDB Endowment, Vol. \vldbvolume, No. \vldbissue\ %
ISSN 2150-8097. \\
\href{https://doi.org/\vldbdoi}{doi:\vldbdoi} \\
}\addtocounter{footnote}{-1}\endgroup

\ifdefempty{\vldbavailabilityurl}{}{
\vspace{.3cm}
\begingroup\small\noindent\raggedright\textbf{PVLDB Artifact Availability:}\\
The source code, data, and/or other artifacts have been made available at \url{\vldbavailabilityurl}.
\endgroup
}
\section{Introduction}\label{sec:intro}

\begin{figure*}[htb!]
\begin{center}
\includegraphics[scale=0.77]{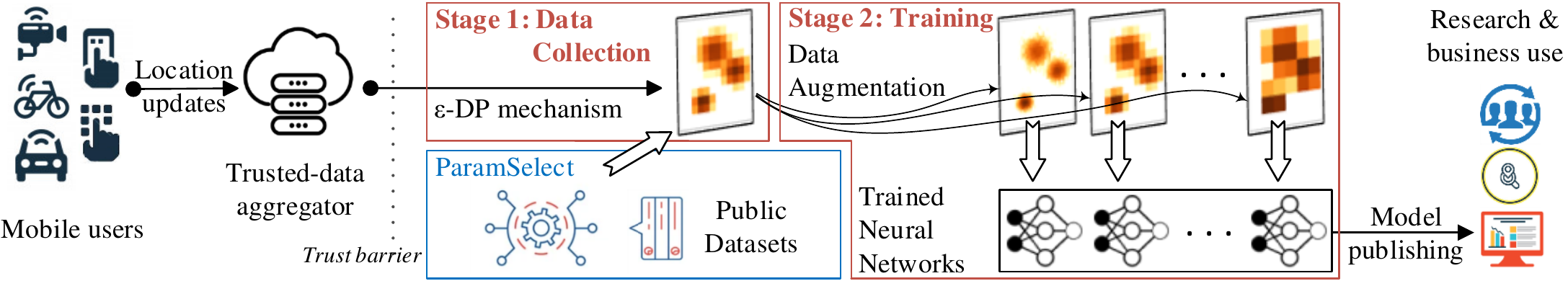}
\end{center}
\caption{Spatial Neural Histogram System}
\label{fig:intro}
\end{figure*}

Mobile apps collect large amounts of individual location data that are beneficial to optimizing traffic, studying disease spread, or improving point-of-interest (POI) placement. When using such data, preserving location privacy is essential, since even aggregate statistics can leak details about individual whereabouts. Most existing solutions to protect location privacy publish a noisy version of
the dataset, transformed according to differential privacy (DP) \cite{dwork2014algorithmic}, the de-facto standard for releasing statistical data. 
The goal of DP mechanisms is to ensure privacy while keeping the answers to queries as accurate as possible. For spatial data, {\em range} queries are the most popular query type, since they can be used as building blocks in most processing tasks. A DP-compliant \textit{representation} of a spatial dataset is created by partitioning the data domain into bins, and then publishing a {\em histogram} with the noisy count of points that fall within each bin. \textit{Domain partitioning} is adopted by nearly all existing algorithms \cite{li2014data, qardaji2013differentially, zhang2016privtree, hay2009boosting, xiao2012dpcube, cormode2012differentially}, e.g., uniform and adaptive grids~\cite{qardaji2013differentially} or hierarchical partitioning~\cite{zhang2016privtree,cormode2012differentially}.

At query time, the noisy histogram is used to compute answers, by considering the counts in all bins that overlap the query. When a query partially overlaps with a bin, the \textit{uniformity assumption} is used to estimate what fraction of the bin's count should be added to the answer. Since DP mechanisms  release only the (noisy) count for each bin, it is assumed that data points are distributed uniformly within the partition, hence the estimate is calculated as the product of the bin count and the ratio of the overlapping area to the total area of the bin. This is often a poor estimate, since location datasets tend to be highly skewed in space (e.g., a shopping mall in a suburb increases mobile user density in an otherwise sparse region). {\em Uniformity error} is a major cause of inaccuracy for existing work on DP release of spatial data.

To address the limitation of existing partitioning-based approaches and reduce query error, we propose a paradigm shift towards a \textit{learned representation} of data. Learning a model to represent accurately the data has gained significant traction in recent years. In the non-private setting, approximate query processing models have been shown to capture data distribution well \cite{ma2019dbest, hilprecht2019deepdb, zeighami2021neurodb}. These models can represent the data compactly, and query answering at runtime is often faster than conventional search algorithms. However, incorporating privacy into such approaches is non-trivial. Recent attempts~\cite{zhang2017privbayes,mckenna2019graphical} propose the use of learned models to answer queries in non-spatial domains (e.g., categorical data). While these approaches perform well in the case of categorical data, they cannot model the intrinsic properties of location datasets, which exhibit both high skewness, as well as strong correlation among regions with similar designations. For instance, two busy downtown areas exhibit similar density patterns, while large regions in between are very sparse. Even at a smaller scale, within a single city, one can identify core community center areas that are very dense, connected by regions with much lower density (e.g., density changes from bars and restaurants to surrounding residential areas). Models with strong representational power in the continuous domain are necessary to learn such patterns.

Meanwhile, training complex models while preserving differential privacy is difficult. In the case of neural networks, existing techniques~\cite{abadi2016deep} utilize \textit{gradient perturbation} to train differentially private models. However, the {\em sensitivity} of this process, defined as the influence a single input record may have on the output (see Section~\ref{sec:prelim} for a formal definition), is high. DP-added noise is proportional to sensitivity, and as a result meaningful information encoded in the gradients is obliterated. For such modeling choices, the learning process has to be carefully crafted by factoring in the unique properties of spatial data, or else the accuracy will deteriorate.

We propose Spatial Neural Histograms (SNH), a neural network system specifically designed to answer differentially private spatial range queries. SNH models answering range queries as \textit{function approximation} tasks, where we learn a function approximator that takes as input a spatial range and outputs the number of points that fall within that range. We train neural networks to model this function while ensuring privacy of the data in the training process. 

Training SNH consists of two stages (Figure~\ref{fig:intro}): the first one perturbs training query answers according to DP, while the second one trains neural networks from noisy answers.
The \textit{training data collection} mechanism in the first stage ensures that sensitivity remains low, such that the signal-to-noise ratio is good. In the second stage, a \textit{data augmentation} step increases the amount of training samples to boost learning accuracy. 
We employ a {\em supervised learning} training process with a carefully selected set of training samples comprising of spatial ranges and their answers. SNH learns from a set of training queries {\em at varying granularity and placement} in order to capture subtle correlations present within the data. 
Finally, system parameters of SNH are tuned with an extensive private parameter tuning step (ParamSelect), which is necessary to evaluate rigorously the privacy-accuracy tradeoff. Parameter selection is performed entirely on publicly available data, without the need to consume valuable privacy budget. 

The fully trained SNH can then be released publicly and only requires a single \textit{forward pass} to answer a query, making it highly efficient at runtime. SNH is able to learn complex density variation patterns that are specific to spatial datasets, and reduces the negative impact of the uniformity assumption when answering range queries, significantly boosting accuracy.


Intuitively, when answering test queries (i.e., at runtime), SNH combines evidence from {\em multiple} training queries from distinct regions and at several granularities. In fact, gradient computation during training can be seen as a novel means of aggregating information across different regions. Learning of underlying patterns from noisy data leads to a significantly more accurate answer. In contrast, existing approaches are limited to using {\em noisy local} information only (i.e., within a single bin). When the noise introduced by differential privacy or the error caused by the uniformity assumption are large for a particular bin, the spatial characteristics of that bin are overwhelmed by the noise and error. As our experiments show, SNH outperforms all the state-of-the-art solutions: PrivTree \cite{zhang2016privtree}, Uniform
Grid (UG) \cite{qardaji2013differentially}, Adaptive Grid (AG) \cite{qardaji2013differentially} and Data and Workload aware algorithm (DAWA) \cite{li2014data}.

Our specific contributions are:
    \begin{itemize}
        \item We formulate the problem of answering spatial range count queries as a function approximation task (Sec.~\ref{sec:prelim});
        \item We propose a novel system that leverages neural networks to represent spatial datasets; the models are carefully designed to accurately capture location-specific density and correlation patterns at multiple granularities (Sec.~\ref{sec:snh},~\ref{sec:snh_details});
        \item We propose a comprehensive framework for tuning system parameters on {\em public} data,  without the need to spend valuable privacy budget. (Sec.\ref{sec:hyperparams});
        \item We conduct an extensive experimental evaluation on a broad array of public and private real-world location datasets with heterogeneous properties (Sec.\ref{sec:exp}); results show that SNH's accuracy considerably outperforms the state-of-the-art.
    \end{itemize}

The rest of the paper is organized as follows:  Section~\ref{sec:prelim} introduces necessary background and definitions. We provide an overview of SNH in Section~\ref{sec:snh} followed by a detailed technical description in Section~\ref{sec:snh_details}. Section~\ref{sec:hyperparams} explores the system parameter selection process, while Section~\ref{sec:exp} presents the experimental evaluation results. We survey  related work in Section~\ref{sec:rel_works} and conclude in Section~\ref{sec:conclusion}.

\section{Preliminaries}\label{sec:prelim}
\subsection{Differential Privacy}
$\varepsilon$-indistinguishability~\cite{dwork2014algorithmic} provides a rigorous privacy framework with formal protection guarantees. Given a positive real number $\varepsilon$, called {\em privacy budget}, a randomized mechanism $\mathscr{M}$ satisfies $\varepsilon$-differential privacy iff for all datasets $D$ and $D^\prime$ differing in at most one element, and for all $E \subseteq$ Range($\mathscr{M}$)
\begin{equation}
Pr[\mathscr{M}(D) \in E] \leq e^\varepsilon Pr[\mathscr{M}(D^\prime) \in E]
\end{equation}
The amount of protection provided by DP increases as $\varepsilon$ approaches $0$. To achieve $\varepsilon$-DP, the result obtained by evaluating a function (e.g., a query) $f$ on the input data must be perturbed by adding random noise. The {\em sensitivity} of $f$, denoted $Z_f$, is the maximum amount the value of $f$ can change when adding or removing a single individual's records from the data. The $\varepsilon$-DP guarantee can be achieved by adding random noise derived from the Laplace distribution $\text{Lap}(Z_f/\varepsilon)$. For a query $f:D \rightarrow \mathbb{R}$, the Laplace mechanism $\mathscr{M}$ returns $f(D) + \text{Lap}(Z_f/\varepsilon)$, where Lap$(Z_f/\varepsilon)$ is a sample drawn from the probability density function Lap$(x|(Z_f/\varepsilon)) = (\varepsilon/2Z_f) \text{exp}(-|x|\varepsilon/Z_f)$ \cite{dwork2014algorithmic}. The {\em composability} property of DP helps quantify the amount of privacy attained when multiple functions are evaluated on the data. 


\subsection{Problem Definition}
Consider a database $D$ that covers a spatial region $SR\subseteq \mathbb{R}^2$, and contains $n$ records each describing an individual's geo-coordinate. Given a privacy budget $\varepsilon$, the problem studied in this paper is to return the answer to an unbounded number of spatial range count queries (RCQs). An RCQ consists of a spatial range predicate and its answer is the number of records in $D$ that satisfy the range predicate. We consider spatial range queries that are axis-parallel and square-shaped, defined by their bottom-left corner $c$ (where $c$ is a vector in $SR$), and their side length $r$. An RCQ, $q$, is then defined by the pair $q=(c, r)$. We say $r$ is the query size and $c$ is its location. For a database $D$, the answer to the RCQ $q=(c, r)$ can be written as a function $f(q)=|\{p|p \in D, c[i]\leq p[i]< c[i]+r, \forall i\in \{\text{lat., lon.}\}\}|$, where $z[\text{lat.}]$ and $z[\text{lon.}]$ denote the latitude and longitude of a point $z$, respectively. We assume RCQs follow a distribution $\mathcal{Q}$ and for any RCQ $q$, we measure the utility of its estimated answer, $y$, using the \textit{relative error metric}, defined as $\Delta(y, f(q))=\frac{|y-f(q)|}{\max\{f(q), \psi\}}$, where $\psi$ is a smoothing factor necessary to avoid division by zero.

The typical way to solve the problem of answering an unbounded number of RCQs is to design an $\varepsilon$-DP mechanism $\mathscr{M}$ and a function $\hat{f}$ such that {\em (1)} $\mathscr{M}$ takes as an input the database $D$ and outputs a differentially private representation of the data, $\theta$; and {\em (2)} the function $\hat{f}(q;\theta)$ takes the representation $\theta$, together with any input query $q$, and outputs an estimate of $f(q)$. 
In practice, $\mathscr{M}$ is used exactly once to generate the representation $\theta$. Given such a representation, $\hat{f}(q;\theta)$ answers any RCQ, $q$, without further access to the database. For instance, in \cite{qardaji2013differentially}, $\mathscr{M}$ is a mechanism that outputs noisy counts of cells of a 2-dimensional grid overlaid on $D$. Then, to answer an RCQ $q$, $\hat{f}(q;\theta)$ takes the noisy grid, $\theta$, and the RCQ, $q$, as inputs and returns an estimate of $f(q)$ using the grid. The objective is to design $\mathscr{M}$ and $\hat{f}$ such that the relative error between $\hat{f}(q;\theta)$ and $f(q)$ is minimized, that is, to minimize $E_{\theta\sim\mathscr{M}}E_{q\sim\mathcal{Q}}[\Delta(\hat{f}(q;\theta), f(q))]$.

As motivated in Sec.~\ref{sec:intro}, we \textit{learn} the data representation used to answer the queries. We let $\hat{f}$ be a function approximator
and define $\mathscr{M}$ to be a mechanism that learns its parameters. The learning objective of $\mathscr{M}$ is to find a $\theta$ such that $\hat{f}(q;\theta)$ closely mimics $f(q)$ for different RCQs, $q$. In this approach, the representation of the data, $\theta$, is the set of learned parameters of a function approximator. Mechanism $\mathscr{M}$ outputs a representation $\theta$, and any RCQ, $q$, is answered by evaluating the function $\hat{f}(q;\theta)$. However, $\mathscr{M}$ is now defined as a learning algorithm and $\hat{f}$ as a function approximator. The problem studied in this paper is formally defined as follows:

\begin{problem}\label{prob:learning_queries}
Given a privacy budget $\varepsilon$, design a function approximator, $\hat{f}$, (let the set of possible parameters of $\hat{f}$ be $\Theta$) and an $\varepsilon$-DP learning algorithm, $\mathscr{M}$, where the objective of $\mathscr{M}$ is to find

$$
\arg\min_{\theta \in \Theta} E_{q\in\mathcal{Q}}[\Delta(\hat{f}(q;\theta), f(q))]
$$
\end{problem}

\section{Spatial Neural Histograms (SNH)}\label{sec:snh}
Our goal is to utilize  models with the ability to learn complex patterns within the data in order to answer RCQs accurately. We employ neural networks as the function approximator $\hat{f}$, due to their ability to learn complex patterns effectively. 

Prior work~\cite{abadi2016deep} introduced the differentially private stochastic gradient descent (DP-SGD) approach to privately train a neural network. Thus, a seemingly straightforward solution to Problem~\ref{prob:learning_queries} is to use a simple fully connected neural network and learn its parameters with DP-SGD. In Sec.~\ref{sec:dp_sgd}, we discuss this trivial approach and outline the limitations of using DP-SGD in our problem setting, which leads to poor accuracy. Next, in Sec.\ref{sec:private_gradient}, we discuss how the training process can be improved to achieve good accuracy. In  Sec.\ref{sec:snh_overview} we provide an overview of our proposed Spatial Neural Histogram (SNH) solution. Table \ref{tbl:notations} summarizes the notations.

\begin{table}
\begin{center}
\begin{tabularx}{0.95\linewidth}{l|X}
	\textbf{Notation}  & \textbf{Definition}   \\\hline
	$\varepsilon$ & DP Privacy Budget \\\hline
	$\mathcal{Q}$, $Q_W$ & Query distribution and workload query set \\\hline
	$Q_D, Y_D$ & Data collection query set and its answers \\\hline
	$Q_A, Y_A$ & Augmented query set and its answers \\\hline
	$R$, $k$ & Set and number of query sizes for training \\\hline
	$l$, $u$ & Lower and upper bound on query sizes \\\hline
	$f(q)$ & Count of records in $D$ that fall in $q$ \\\hline
	$\hat{f}(q;\theta)$ & Count of records in $q$ estimated from $\theta$  \\\hline
	$\bar{f}(q)$ & $f(q)+Lap(\frac{1}{\varepsilon})$  \\\hline
	$\rho$ & Grid granularity 	\\\hline
	$C$ & Set of bottom-left corners of grid cells \\\hline
	$\psi$ & Smoothing factor in relative error\\\hline
	$\Phi$,$\phi$ & ParamSelect Model, Dataset features\\\hline
	$\mathcal{D}$,$D^*$ & Public ParamSelect training and inference datasets\\\hline
	$\pi_\alpha(D, \varepsilon)$  & Function denoting best value of system parameter $\alpha$ for dataset $D$ and budget $\varepsilon$ \\\hline
	$\hat{\pi}_\alpha(D, \varepsilon)$  & Empirical estimate of $\pi_\alpha(D, \varepsilon)$ \\

\end{tabularx}
\end{center}
\caption{Summary of Notations}
\label{tbl:notations}
\end{table}

\subsection{Baseline Solution using DP-SGD}\label{sec:dp_sgd}
We define $\hat{f}(.;\theta)$ to be a fully connected neural network with parameter set $\theta$. $\hat{f}(q;\theta)$ takes an RCQ $q=(c, r)$ as input 
and outputs a single real number. The goal is to learn the parameters of such a neural network.

\noindent\textbf{Learning Setup}. We train the neural network so that for an RCQ $q$, its output $\hat{f}(q;\theta)$ is similar to $f(q)$. This can be formulated as a typical supervised learning problem. A training set, $T$, is created, consisting of $(q, f(q))$ pairs, where $q$ is the input to the neural network and $f(q)$ is the training label for the input $q$ (we call RCQs in the training set \textit{training RCQs}). To create the training set, similar to \cite{li2014data, mckenna2018optimizing}, we assume we have access to a set of \textit{workload RCQs}, $Q_W$, that resembles RCQs a query issuer would ask (e.g., are sampled from $\mathcal{Q}$ or a similar distribution) and is assumed to be public. Thus, we can define our training set $T$ to be $\{(q, f(q))|q\in Q_W\}$. Using this training set, we define the training loss as

\begin{align}\label{eq:loss}
 \mathcal{L}=\sum_{q\in Q_W}(\hat{f}(q;\theta)-f(q))^2   
\end{align}

In a non-private setting, a model can be learned by directly optimizing Eq.~\eqref{eq:loss} using a gradient descent approach, and a successfully learned model will be able to answer any new RCQ $q$ similar to the ground truth $f(q)$. However, in a private setting, every gradient calculation accesses the database (since the labels for each training sample are calculated from the record attributes), and thus has to be done in a privacy-preserving manner. Each  gradient calculation consumes privacy budget, and since it is done at every iteration of learning, privacy consumption is very high. 

\noindent\textbf{Incorporating Privacy}. DP-SGD \cite{abadi2016deep} incorporates differential privacy for training neural networks. It modifies traditional SGD by clipping the gradient values and obfuscating them with Gaussian noise. Specifically, in each iteration: (1) a subset, $S$, of the training set is sampled; (2) for each sample, $s=(x, y)\in S$, the gradient $g_s=\nabla_\theta(\hat{f}(x;\theta)-y)^2$ is computed, and clipped (i.e., truncated) to a maximum $\ell_2$-norm of $B$ as $\bar{g}_s=\min(\lVert g_s\rVert_2, B)\frac{g_s}{\lVert g_s\rVert_2}$; (3) the average clipped gradient value for samples in $S$ is obfuscated with Gaussian noise as 
\begin{align}\label{eq:private_gradient}
    g=\sum_{s\in S}(\bar{g}_s)+\mathscr{N}(0, \sigma^2B^2)
\end{align} (4) the parameters of the neural network are updated in the direction opposite to $g$. 


\noindent\textbf{DP-SGD Challenges}. Applying DP-SGD to our problem setting is not straightforward. In fact, a straightforward application of DP-SGD does not satisfy the privacy requirement of Problem~\ref{prob:learning_queries}. In our problem setting, the training set is created by querying $D$ to obtain the training labels, and our goal is to ensure the privacy of records in $D$. On the other hand, DP-SGD considers the training set itself, to be the dataset whose privacy needs to be secured. This changes the sensitivity analysis of DP-SGD. In our setting, to compute the sensitivity of the gradient sum $\sum_{s\in S}(\bar{g}_s)$ in step (3) of DP-SGD, we have to consider the worst-case effect the presence or absence of a single geo-coordinate record $p$ can have on the sum (as opposed to the worst-case effect of the presence or absence of a single training sample). Since removing $p$ can potentially affect every $\bar{g}_s$ for all $s \in S$, the sensitivity of the gradient sum is $|2S|\times B$. Thus, Gaussian noise of $\mathscr{N}(0, \sigma^2 4|S|^2 B^2)$ must be added to the gradient sum to achieve DP  (compare with noise in step (3) above).  After this adjustment, the per-iteration and total privacy consumption of DP-SGD is amplified, impairing learning. We experimentally observed that, for any reasonable privacy budget, the training loss does not improve at all during training, due to the large amount of noise being added.

\subsection{A different learning paradigm for RCQs}\label{sec:private_gradient}
We propose a new learning paradigm for answering RCQs that addresses the shortcomings of DP-SGD. In this section, we present {\em three design principles (P1-P3)} we follow when training neural networks to answer RCQs. These principles are then used in Sec.~\ref{sec:snh_overview} to build our solution. 


\noindent\textbf{P1: Separation of noise addition from training}. The main reason DP-SGD fails in our problem setting is that too much noise needs to be added when calculating gradients privately. Recall that DP-SGD uses the quantity $g$, defined in Eq.~\eqref{eq:private_gradient}, as the differentially private estimate of the gradient of the loss function. Here, we investigate the private gradient computation in more details to provide an alternative method to calculate the gradient with differential privacy. 
Recall that the goal is to obtain the gradient of the loss function, $\mathcal{L}$, defined in Eq.~\eqref{eq:loss} with respect to the model parameters. We thus differentiate $\mathcal{L}$ and obtain: 
\bgroup
\def\arraystretch{0.5}
\begin{align}\label{eq:grad_diff}
    \nabla_\theta\mathcal{L}=\sum_{q\in Q_W}2\times(\underbrace{\hat{f}(q;\theta)}_{\text{\begin{tabular}{c}
         data  \\
         indep.
    \end{tabular}}}-\underbrace{f(q)}_{\text{\text{\begin{tabular}{c}
         data  \\
         dep.
    \end{tabular}}}})\times\underbrace{\nabla\hat{f}(q;\theta)}_{\text{\begin{tabular}{c}
         data  \\
         indep.
    \end{tabular}}}
\end{align}
\egroup
In Eq.~\eqref{eq:grad_diff}, only $f(q)$ accesses the database. This is because the training RCQs in $Q_W$ (i.e., the inputs to the neural network), are created independently of the database. The data dependent term requires computing private answers to $f(q)$ for an RCQ $q$, hence must consume budget, while the data-independent terms can be calculated without spending any privacy budget. This decomposition of the gradient into data dependent and independent terms is possible because, different from typical machine learning settings, the differential privacy is defined with respect to the database $D$ and not the training set (as discussed in Sec.~\ref{sec:dp_sgd}). 

Thus, instead of using $g$, as defined in Eq.~\eqref{eq:private_gradient}, as the differentially private estimate of the gradient (where the gradients are clipped and noise is added to the clipped gradients), we calculate a differentially private value of  the training label $f(q)$, called $\bar{f}(q)$, by adding noise to the label (define $\bar{f}(q)=f(q)+\text{Lap}(1/\varepsilon)$) and calculate the gradient from that. The differentially private estimate of the gradient is then 
\bgroup
\def\arraystretch{0.5}
\begin{align}\label{eq:grad_diff_new}
    g=\sum_{q\in Q_W}2\times(\hat{f}(q;\theta)-\bar{f}(q))\times\nabla\hat{f}(q;\theta)
\end{align}
\egroup
A crucial benefit is that $\bar{f}(q)$, does not change over successive learning iterations. That is, the differentially private value $\bar{f}(q)$ can be computed once and used for all training iterations. This motivates our first design principle of separating noise addition and training. This way, training becomes a two step process: first, for all $q\in Q_W$, we calculate the differentially private  training label $\bar{f}(q)$. We call this step \textit{data collection}. Then, we use a training set consisting of pairs $(q, \bar{f}(q))$ for all $q\in Q_W$ for training. Since DP-compliant data measurements are obtained, all future operations that use as input these measurements are also $\varepsilon$-differentially private according to the \textit{post-processing property} of differential privacy~\cite{dwork2014algorithmic}. Thus, the training process is done as in a non-private setting, where a conventional SGD algorithm can be applied (i.e., we need not add noise to gradients), and differential privacy is still satisfied.

\noindent\textbf{P2: Spatial data augmentation through partitioning}. In the approach discussed above, the only time privacy budget is spent is when calculating $\bar{f}(q)$ in Eq.~\eqref{eq:grad_diff_new}, where $\bar{f}(q)$ is calculated for the $|Q_W|$ different $q\in Q_W$. Answering any $q\in Q_W$ consumes privacy budget $\varepsilon$ (since noise $Lap(\frac{1}{\varepsilon})$ is added, and the sensitivity of the RCQ set is one). If the spatial extent of RCQs in $Q_W$ overlap (i.e., an individual's record could be accessed by multiple RCQs), \textit{sequential composition} theorems would apply when accounting for the total privacy budget consumed. Thus, the more training samples used, the more privacy budget needs to be spent. Meanwhile, overlapping training queries are needed for accurate training. This is because a neural network needs to see queries of multiple query sizes for the same location during training to be able to answer new unseen queries with different sizes. 

To resolve the above dilemma, we propose \textit{spatial data augmentation} through partitioning. First, instead of accessing the database queries in $Q_W$, we use a different \textit{data collection query set}, $Q_D$, chosen such that the RCQs in $Q_D$ don't overlap (i.e., a partitioning of the space). This ensures \textit{parallel composition} can be used for privacy accounting, instead of sequential composition, which avoids sensitivity escalation over multiple RCQs in $Q_D$. Only RCQs in $Q_D$ access the data. Subsequently, through data augmentation, RCQs in $Q_D$ are aggregated or split to generate more training samples. For instance, multiple RCQs next to each other can be aggregated to answer an RCQ with a larger query size. Data augmentation allows us to use training samples with multiple query sizes over the same location without spending extra privacy budget collecting them.

\noindent\textbf{P3: Learning at multiple granularities}. We employ in our solution  multiple models that learn at different granularities, each designed to answer RCQs of a specific size. Intuitively, it is more difficult for a model to learn patterns when both query size and locations change. Using multiple models allows each model to learn the patterns relevant to the granularity they operate on.

\begin{figure}[tb!]
\begin{center}
\includegraphics[width=\columnwidth]{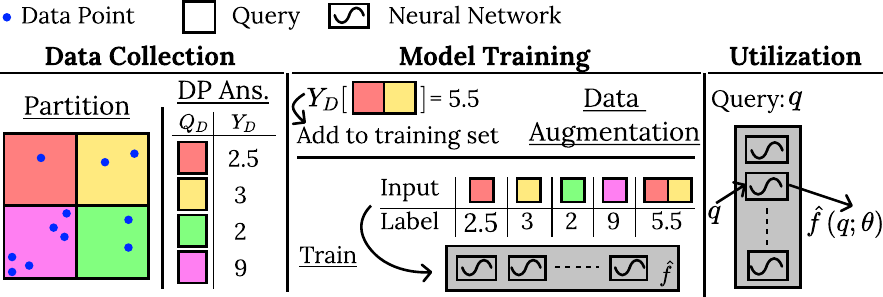}
\caption{SHN Overview}\label{fig:overview}
\end{center}
\end{figure}

\subsection{Proposed approach: SNH}\label{sec:snh_overview}
Guided by the principles from Sec.~\ref{sec:private_gradient}, we introduce the Spatial Neural Histograms (SNH) design illustrated in Figure~\ref{fig:overview}. It consists of three steps: (1) Data collection, (2) Model Training, and (3) Model Utilization. We provide a summary of each step below, and defer details until Sec.~\ref{sec:snh_details}. 

\noindent\textbf{Data Collection}. This step partitions the space into non-overlapping RCQs that are directly answered with DP-added noise. The output of this step is a \textit{data collection query set}, $Q_D$, and a set $Y_D$ which consists of the differentially private answers to RCQs in $Q_D$. This is the only step in SNH that accesses the database. In Fig.~\ref{fig:overview} for example,  the query space is partitioned into four RCQs, and a differentially private answer is computed for each. 
 
\noindent\textbf{Training}. 
Our training process consists of two stages. First, we use \textit{spatial data augmentation} to create more training samples based on $Q_D$. An example is shown in Fig.~\ref{fig:overview}, where an RCQ covering both the red and yellow squares is not present in the set $Q_D$, but it is obtained by aggregating its composing sub-queries (both in $Q_D$). Second, the augmented training set is used to train a function approximator $\hat{f}$ that captures $f$ well. $\hat{f}$ consists of a set of neural networks, each trained to answer different query sizes. 



\noindent\textbf{Model Utilization}. This step decides how any previously unseen RCQ can be answered using the learned function approximator, and how different neural networks are utilized to answer an RCQ.

\section{Technical Details}\label{sec:snh_details}
In this section, we discuss details of the three SNH steps.
\begin{figure}
    \centering
    \includegraphics[width=\columnwidth]{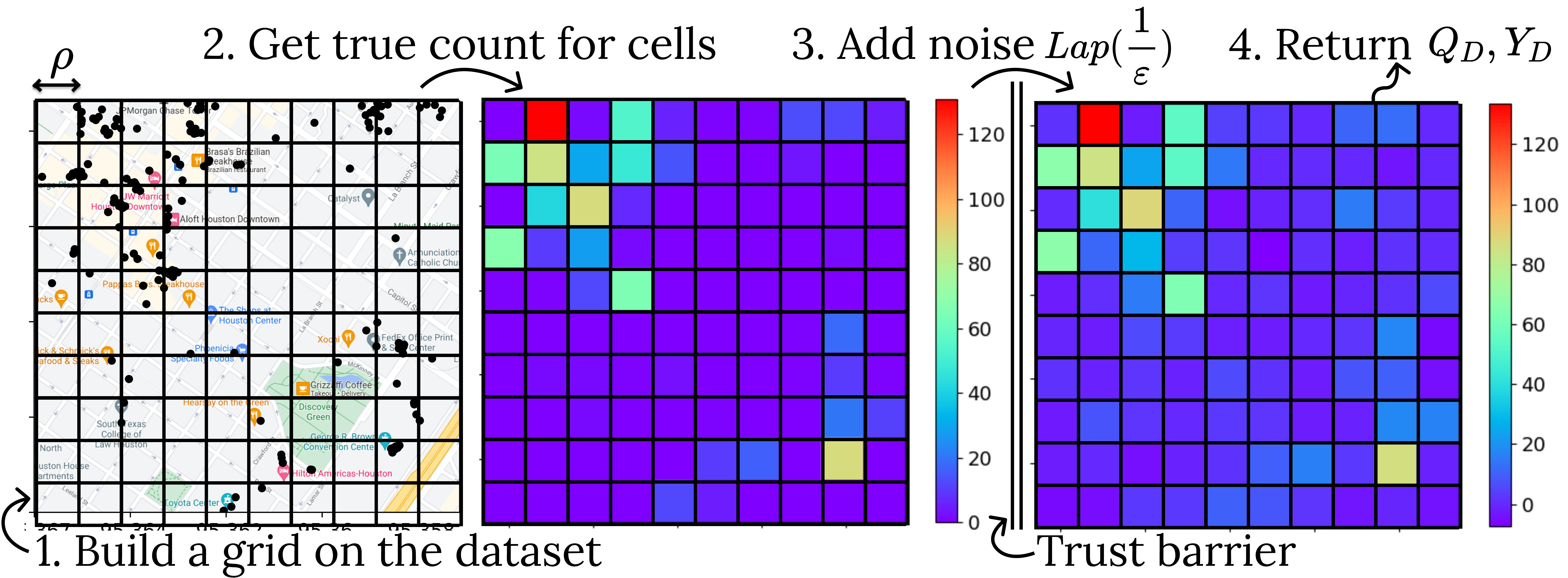}
    \caption{Data Collection}
    \label{fig:data_collection}
\end{figure}

\subsection{Step 1: Data Collection}
This step creates a partitioning of the space into non-overlapping bins, and computes for each bin a  differentially private answer. 
We opt for a simple equi-width grid of cell width $\rho$ as our partitioning method. As illustrated in Fig.~\ref{fig:data_collection}, {\em (1)} we overlay a grid on top of the data domain; {\em (2)} we calculate the true count for each cell in the grid, and {\em (3)} we add noise sampled from $Lap(\frac{1}{\varepsilon})$ to each cell count. We represent a cell by the coordinates of its bottom left corner, $c$, so that getting the count of records in each cell is an RCQ, $q=(c, \rho)$. Let $C$ be the set of bottom left coordinates of all the cells in the grid. Furthermore, recall that for a query $q$, $\bar{f}(q)=f(q)+Lap(\frac{1}{\varepsilon})$. Thus, the data collection query set is defined as $Q_D=\{(c, \rho), c\in C\}$, and their answers are the set $Y_D=\{\bar{f}(c, \rho), c\in C \}$. We use $Y_D[c]$ to refer to the answer for the query located at $c$ in $Y_D$. The output of the data collection step consists of sets $Q_D$ and $Y_D$.

Even though more complex partitioning structures have been used previosuly for privately answering RCQs~\cite{qardaji2013differentially,zhang2016privtree}, we chose a simple regular grid, for two reasons. First, our focus is on a novel neural database approach to answering RCQs, which can be used in conjunction with {\em any} partitioning type -- using a simple grid allows us to isolate the benefits of the neural approach. Second, using more complex structures in the data collection step may increase the impact of uniformity error, which we attempt to suppress through our approach. The neural learning step captures density variations well, and conducting more complex DP-compliant operations in the data collection step can have a negative effect on overall accuracy. 
In our experiments, we observed significant improvements in accuracy with the simple grid approach. While it may be possible to improve the accuracy of SNH by using more advanced data collection methods, we leave that study for future work.

The primary challenge with respect to SNH data collection is choosing the value of $\rho$, which is well-understood to impact the privacy-accuracy trade-off \cite{qardaji2013differentially}. We address this thoroughly in Sec.~\ref{sec:hyperparams} and present a method to determine the best granularity of the grid. 

\begin{figure}
    \centering
    \includegraphics[width=\columnwidth]{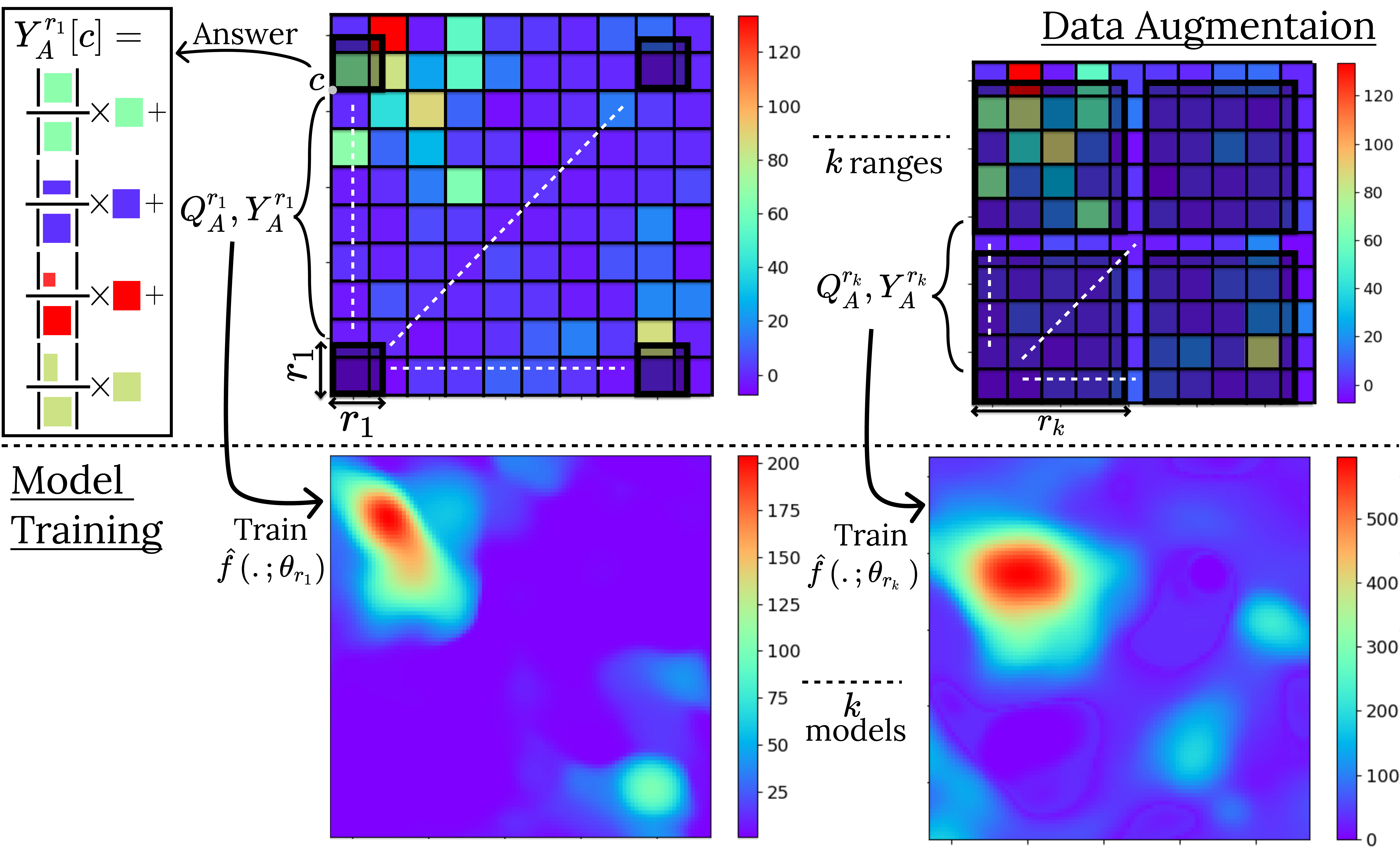}
    \caption{Model Training}
    \label{fig:training}
\end{figure}
\subsection{Step 2: SNH Training} Given the query set $Q_D$ and its sanitized answers, we can now perform any operation on this set without worrying about privacy leakage. As discussed in Sec.~\ref{sec:snh_overview}, we first perform a data augmentation step using $Q_D$ to create an augmented training set $Q_A$. Then, $Q_A$ is used for training our function approximator. 

\noindent\textbf{Data Augmentation} is a common  machine learning technique to increase the number of samples for training based on the existing (often limited) available samples \cite{zhong2020random, krizhevsky2012imagenet}. Here, we propose spatial data augmentation for learning to answer RCQs. Our data collection query set $Q_D$ contains RCQs for different data domain regions, but all with same query size $\rho$. Thus, a model learned using $Q_D$ as a training set will have no information about queries of different sizes, hence will be unable to answer them accurately. 

As such, we use the set $Q_D$ and answers $Y_D$ to estimate the answers for queries at the same locations but of {\em other} sizes. Specifically, consider a query location $c\in C$ and a query size $r$, $r\neq \rho$. We estimate the answer for RCQ $q=(c, r)$ as $\sum_{c'\in C}\frac{|(c, r)\cap (c', \rho)|}{\rho^2}\times Y_D[c]$, where $|(c, r)\cap (c', \rho)|$ is the overlapping area of RCQs $(c, r)$ and $(c',r)$. In this estimate, noisy counts of cells in $Q_D$ fully covered by $q$ are added as-is (since $|(c, r)\cap (c', \rho)|=\rho^2$), whereas fractional counts for partially-covered cells are estimated using the uniformity assumption. 
Fig.~\ref{fig:training} shows how we perform data augmentation for a query $(c, r_1)$ with size $r_1$ at location $c$. Also observe that the bottom-left corners of all queries in the augmented query set are aligned with the grid.

We repeat this procedure for $k$ different query sizes to generate sufficient training data. To ensure coverage for all expected query sizes, we define the set of $k$ sizes to be uniformly spaced. Specifically, assuming the test RCQs have size between $l$ and $u$, we define the set $R$ as the set of $k$ uniformly spaced values between $l$ and $u$, and we create an augmented training set for each query size in $R$. This procedure is shown in Alg.\ref{alg:train}. We define $Q_A^r$ for $r\in R$ to be the set of RCQs located at $C$ but with query size $r$, that is $Q_A^r=\{(c, r), c\in C\}$, and define $Y_A^r$ to be the set of the estimates for queries in $Q_A^r$ obtained from $Q_D$ and $Y_D$. The output of Alg.~\ref{alg:train} is the augmented training set containing training samples for different query sizes. Note that, as seen in the definition above,  $Q_A^r$, for any $r$, only contains queries whose bottom-left corner is aligned with the grid used for data collection to minimize the use of the uniformity assumption.

\begin{algorithm}[t]
\begin{algorithmic}[1]
\Require Query set $Q_D$ with answers $Y_D$, $k$ query sizes
\Ensure Augmented training set $Q_A$ with labels $Y_A$
\State $R \leftarrow \{l+\frac{(u-l)}{k}\times (i+\frac{1}{2}), \forall i,  0\leq i< k \}$ \label{alg:train:ranges}
\ForAll{$r\in R$}  
    \State $Q_A^r, Y_A^r\leftarrow \emptyset$
    \For{ $(c, \rho)\in Q_D$}
    \State $Q_A^r.append((c, r))$
    \State $Y_A^r[c]\leftarrow \sum_{(c', \rho)\in Q_D}\frac{|(c, r)\cap (c', \rho)|}{\rho^2}\times Y_D[c']$\label{alg:train:get_sample}
    \EndFor
\EndFor
\State \Return $Q_A, Y_A\leftarrow \{Q_A^r, \forall r\in R \}, \{Y_A^r, \forall r\in R \}$
\end{algorithmic}
\caption{Spatial data augmentation}\label{alg:train}
\end{algorithm}

\noindent\textbf{Model architecture}. We find that using multiple neural networks, each trained for a specific query size, performs better than using a single neural network to answer queries of all sizes. Thus, we train $k$ different neural networks, one for each $r\in R$. Meaning that a single neural network trained for query size $r$ can only answer queries of size $r$ (we discuss in Sec.~\ref{sec:utilization} how the neural networks are used to answer other query sizes), accordingly the input dimensionality of each neural network is two, i.e., lat. and lon. of the location of the query. We use $k$ identical fully-connected neural networks (specifics of the network architecture are discussed in Sec.~\ref{sec:exp}).

\noindent\textbf{Loss function and Optimization}. 
We train each of the $k$ neural networks independently. We denote by $Q_A^{r}$ the training set for a neural network $\hat{f}(.;\theta_r)$, trained for query size $r$, and we denote the resulting labels by $Y_A^r$. We use a mean squared error loss function to train the model, but propose two adjustments to allow the neural network to capitalize on the workload information available. First, note that for a query size $r\in R$, $Q_A^r$ is comprised of queries at uniformly spaced intervals, which does not necessarily follow the query distribution $\mathcal{Q}$. However, we can exploit properties of the workload queries, $Q_W$ to tune the model for queries from $\mathcal{Q}$. Specifically, for any $(c, r)\in Q_A^r$, let $w_{(c, r)}=|\{q'\in Q_W, (c, r)\cap q' \neq\emptyset\}|$, that is, $w_{(c, r)}$ is the number of workload queries that overlap a training query. In our loss function, we weight every query $(c, r)$ by $w_{(c, r)}$. This workload-adaptive modifications to the loss function emphasizes the regions that are more popular for a potential query issuer. Second, we are interested in answering queries with low relative error, whereas a mean square loss puts more emphasis on absolute error. Thus, for a training query $(c, r)$, we also weight the sample by $\frac{1}{\max\{Y_A^r[c], \psi\}}$. Putting these together, the loss function optimized for each neural network is 
\begin{align}\label{eq:final_loss}
    \sum_{(c, r)\in Q_A^r} \frac{w_{(c, r)}}{\max\{Y_A^r[c], \psi\}}(\hat{f}(c, \theta_r)-Y_A^r[c])^2
\end{align}

\begin{figure}
    \centering
    \includegraphics[width=\columnwidth]{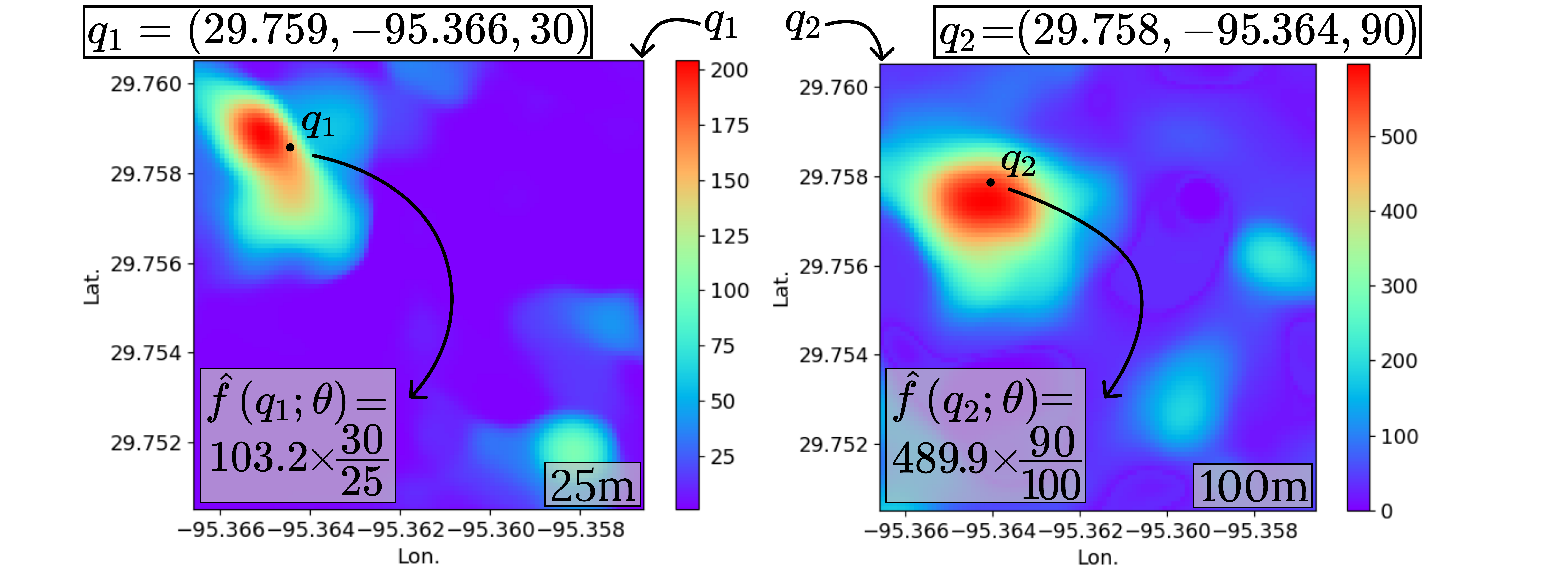}
    \caption{Model utilization for two RCQs}
    \label{fig:utilization}
\end{figure}

\subsection{Model Utilization} \label{sec:utilization}
Once the SNH system consisting of neural histograms at various query sizes is fully trained, unseen queries $q\in Q$ can be answered. To answer a query $(c, r)$ the neural network that is trained to answer queries with size most similar to $r$ is accessed. That is, we find $r^*=\arg\min_{r'\in R}|r-r'|$ and we answer the query using network $\hat{f}(c, \theta_{r^*})$. The output answer is scaled to $r$ according to a uniformity assumption, and the scaled answer is returned, i.e., $\frac{r}{r^*}\hat{f}(c, \theta_{r^*})$. Fig.~\ref{fig:utilization} shows this procedure for two different RCQs. 

Note that, the impact of number of models trained, $k$, can be seen in the ratio $\frac{r}{r^*}$ when answering the queries. The larger $k$ is, the closer this ratio will be to $1$ and thus the uniformity error on the \textit{output} of the neural network will be smaller. However, larger $k$ means more training samples are collected using uniformity assumption. Thus, an interesting trade-off arises when comparing the impact of uniformity assumption \textit{pre-learning} vs. \textit{post-learning}. We empirically observed that the accuracy improves as we increase $k$, that is, the use of uniformity assumption post learning should be minimized. Intuitively, this is caused by how learning counters the effect of errors introduced by uniformity assumption by using learned patterns from the data.

\section{ParamSelect: Optimizing Grid discr- etization and system parameters}\label{sec:hyperparams}
Grid granularity, $\rho$, is a crucial parameter for achieving good performance for DP publishing of spatial data, and selecting it has been the subject of previous work \cite{qardaji2013differentially, hay2016principled}. Discretizing continuous domain geo-coordinates creates uniformity errors, and as a result the granularity of the grid must be carefully tuned to compensate for the effect of discretization. Existing work \cite{qardaji2013differentially, hay2016principled} makes simplifying assumptions to analytically model the impact of grid granularity on the accuracy of answering queries. However, modelling data and query specific factors is difficult and the simplifying assumptions are often not true in practice, as we show in our experiments (see Sec.~\ref{sec:exp:paramselect}). Instead, we learn a model that is able to predict an advantageous grid granularity for the specific dataset, query distribution and privacy budget. In Sec.~\ref{sec:paramselect:rho}, we discuss {\em ParamSelect}, our approach for determining the value of $\rho$. In Sec.~\ref{sec:paramselect:any} we show how {\em ParamSelect}
can be extended to tune other system parameters.


\subsection{ParamSelect for $\rho$}\label{sec:paramselect:rho}
The impact of grid granularity on privacy-accuracy trade-offs when answering queries is well-understood in the literature \cite{qardaji2013differentially}. In SNH, the grid granularity in \textit{data collection} phase impacts the performance as follows. On the one hand, smaller grid cells increase the resolution at which the data are collected, thereby reducing the uniformity error. Learning is also improved, due to more training samples being extracted. On the other hand, creating too fine grids can diminish the signal-to-noise ratio for cells with small counts, since at a given $\varepsilon$ the magnitude of noise added to any cell count is fixed. Moreover, during data augmentation, aggregating multiple cells leads to increase in the total noise variance, since the errors of individual cells are summed. SNH is thus impacted by the cell width in multiple ways, and determining an advantageous cell width, $\rho$, is important to achieving good accuracy.

Meanwhile, capturing an analytical dependence may not be possible, since numerous data, query and modelling factors determine the ideal cell width. 
For instance, 
if data points are concentrated in some area and the queries also fall in the same area, a finer grid can more accurately answer queries for the query distribution (even though signal-to-noise ratio may be poor for parts of the space where queries are not often asked). This factor can be measured only by looking at the actual data and the distribution of queries, and would require spending privacy budget.

Overall, the best value of $\rho$ depends on the privacy budget $\varepsilon$, the size and distribution of points of $D$ and the query distribution $\mathcal{Q}$. Thus, we denote by $\pi(D, \varepsilon)$ a function that, given a query distribution, outputs the ideal grid cell width for a certain database and privacy budget. 
We learn a model, $\Phi$, that approximates $\pi(D, \varepsilon)$. The learning process is similar to any supervised learning task, where for different dataset and privacy budget pairs, $(D, \varepsilon)$, we use the label $\pi(D, \varepsilon)$ to train $\Phi$. The input to the model is $(D, \varepsilon)$ and the training objective is to get the model output, $\Phi(D, \varepsilon)$, to be close to the label $\pi(D, \varepsilon)$.

\noindent\textbf{Feature engineering.}
 Learning a model that takes a raw database $D$ as input is infeasible, due to the high sensitivity of learning with privacy constraints. Instead, we introduce a feature engineering step that, for the dataset $D$, outputs a set of features, $\phi_D$. Model training then replaces $D$ with $\phi_D$. Let the spatial region of $D$ be $SR_{D}$. First, as one of our features, we measure the skewness in the spread of individuals over $SR_{D}$, since this value directly correlates with the expected error induced by using the uniformity assumption. In particular, we (1) discretize $SR_{D}$ using an equi-width partitioning, (2) for each cell, calculate the probability of a point falling into a cell as the count of points in the cell normalized by total number of points in $D$, and (3) take the Shannon's Entropy $h_{D}$ over the probabilities in the flattened grid. However, calculating $h_{D}$ on a private dataset violates differential privacy. Instead, we utilize {\em publicly available} location datasets as an auxiliary source to approximately describe the private data distribution for the same spatial region. We posit that there exist high-level similarities in distribution of people's locations in a city across different private and public datasets for the same spatial regions and thus, the public dataset can be used as a surrogate. Specifically, let $D^*$ be a public dataset covering the same spatial region as $D$. We estimate $h_D$ for a dataset with $h_{D^*}$. We call $D^*$ \textit{public ParamSelect inference dataset}.

Second, we use data-independent features: $\varepsilon$, $\frac{1}{n\times\varepsilon}$ and $\frac{1}{\sqrt{n\times\varepsilon}}$, where the product of $n\times\varepsilon$ accounts for the fact that decreasing the scale of the input dataset and increasing epsilon have equivalent effects on the error. This is also understood as epsilon-scale exchangeability \cite{hay2016principled}. We calculate $\phi_{D, \varepsilon}=(n,\varepsilon,  \frac{1}{n\varepsilon},  \frac{1}{\sqrt{n\varepsilon}}, h_{D^*})$ as the set of features for the dataset $D$ without consuming any privacy budget in the process. Lastly, we remark that for regions where an auxiliary source of information is unavailable, we may still utilize the data-independent features to good effect.


\noindent\textbf{Training Sample Collection}. Generating training samples for $\Phi$ is not straightforward since we do not have an analytical formulation for $\pi(D, \varepsilon)$. While the exact value of $\pi(D, \varepsilon)$ is unknown, an empirical estimate is a good alternative. We run SNH with various grid granularities of \textit{data collection} and return the grid size, $\rho_{D, \varepsilon}$, for which SNH achieves the lowest error. In this way we obtain $\rho_{D, \varepsilon}$ as our training label. Note that the empirically determined value of $\rho_{D, \varepsilon}$ is dependent on---and hence accounts for---the query distribution on which SNH error is measured. Moreover, when $D$ contains sensitive data, obtaining $\rho_{D, \varepsilon}$ would require spending privacy budget. Instead, we generate training records from a set of datasets, $\mathcal{D}$ that have already been publicly released (see Sec.~\ref{sec:exp} for details of public datasets). We call datasets in $\mathcal{D}$ \textit{public ParamSelect training datasets}. 
Putting everything together, our training set is $\{(\phi_{D, \varepsilon}, \rho_{D, \varepsilon}) | \varepsilon\in\mathcal{E}, D\in\mathcal{D}\}$, where $\mathcal{E}$ is the range of different privacy budgets chosen for training.


\begin{algorithm}[t]
\caption{ParamSelect training} \label{alg:paramselect_train}
\begin{algorithmic}[1]
\Require{A set of public datasets $\mathcal{D}$ and privacy budgets $\mathcal{E}$ for training to predict a system parameter $\alpha$ }
\Ensure{Model $\Phi_\alpha$ for system parameter $\alpha$}
\Procedure{$\phi$}{$D, n, \varepsilon$}
\State $h_{D}\leftarrow$ entropy of $D$
\State \Return $(n,\varepsilon, \frac{1}{n\varepsilon},  \frac{1}{\sqrt{n\varepsilon}}, h_{D})$
\EndProcedure
\Procedure{Train\_ParamSelect}{$\mathcal{D}$, $\mathcal{E}$}
\State $T\leftarrow\{(\phi(D, |D|, \varepsilon), \hat{\pi}_{\alpha}(D, \varepsilon)) | \varepsilon\in\mathcal{E}, D\in\mathcal{D}\}$
\State $\Phi_\alpha\leftarrow$ Train model using $T$
 \State \Return $\Phi_\alpha$
\EndProcedure

\end{algorithmic}
\end{algorithm}

\begin{algorithm}[t]
\caption{ParamSelect usage} \label{alg:paramselect}
\begin{algorithmic}[1]
\Require{Spatial extent $SR$ and size $n$ of a sensitive dataset $D$ and privacy budget $\varepsilon$}
\Ensure{System parameter value $\alpha$ for private dataset $D$}
\Procedure{ParamSelect}{$SR$, $n$, $\varepsilon$}
    \State $D^* \leftarrow$ Public dataset with spatial extent $SR$  
  \State $\alpha \leftarrow \Phi_\alpha(\phi(D^*, n, \varepsilon))$
  \State \Return $\alpha$
\EndProcedure
\end{algorithmic}
\end{algorithm}

\noindent\textbf{Predicting Grid Width with ParamSelect.} The training phase of ParamSelect builds a regression model, $\Phi$  using the training set described above. We observe that models from the decision tree family of algorithms perform the best for this task. Once the model is trained, its utilization for any unseen dataset is straightforward and only requires calculating the corresponding features and evaluating $\Phi$ for the dataset. 

\subsection{Generalizing ParamSelect to any system parameter}\label{sec:paramselect:any}
We can easily generalize the approach in Sec.~\ref{sec:paramselect:rho} to any system parameter. Define function $\pi_\alpha(D, \varepsilon)$ that given a query distribution, outputs the best value of $\alpha$ for a certain database and privacy budget. The goal of ParamSelect is to learn a model, using public datasets $\mathcal{D}$, that mimics the function $\pi_\alpha(.)$. 

ParamSelect functionality is summarized in Alg.~\ref{alg:paramselect_train}. First, during a pre-processing step, it defines a feature extraction function $\phi(D, n, \varepsilon)$, that extracts relevant features from the public dataset $D$ with $n$ records, and a privacy budget $\varepsilon$. Second, it creates the training set $\{(\phi(D, |D|, \varepsilon), \hat{\pi}_{\alpha}(D, \varepsilon)), \varepsilon\in\mathcal{E}, D\in\mathcal{D}\}$, where $\hat{\pi}_{\alpha}(D, \varepsilon)$ estimates the value of $\pi_{\alpha}(D, \varepsilon)$ with an empirical search (i.e., by trying different values of $\alpha$ and selecting the one with the highest accuracy), and $\mathcal{D}$ and $\mathcal{E}$ are different public datasets and values of privacy budget, respectively, used to collect training samples. Lastly, it trains a model $\Phi_{\alpha}$ that takes extracted features as an input and outputs a value for $\alpha$ in a supervised learning fashion. 

At model utilization stage (Alg.~\ref{alg:paramselect}), given a sensitive dataset $D$, ParamSelect uses a public dataset $D^*$ that covers the same spatial region, as well as size of $D$, $n$, and privacy budget $\varepsilon$ to extract features $\phi(D^*, n, \varepsilon)$. The predicted system parameter value for $D$ is then $\Phi_{\alpha}(\phi(D^*, n, \varepsilon))$.

\section{Experimental Evaluation}\label{sec:exp}
Section~\ref{subsec:settings} describes the experimental testbed. Section~\ref{subsec:comparison} evaluates SHN in comparison with state-of-the-art approaches. Sections~\ref{subsec:ablation} and~\ref{subsec:hyperparam} provide an ablation study and an evaluation of the impact of varying system parameters, respectively.

\subsection{Experimental Settings}
\label{subsec:settings}
\subsubsection{Datasets} \label{exp:setup:dataset}
We evaluate SNH on both proprietary and public-domain datasets. Each of our experiments uses a private dataset (which needs to be protected when answering RCQs) and auxiliary datasets that are publicly available. The auxiliary datasets are used to determine: (1) workload queries, $Q_W$ (used in SNH training loss function, Eq.~\eqref{eq:final_loss}), (2) public ParamSelect training dataset $\mathcal{D}$ (see Sec.~\ref{sec:paramselect:any}) and (3) public ParamSelect inference dataset $D^*$ (see Sec.~\ref{sec:paramselect:any}). We first describe all the datasets and then specify how they are utilized in our experiments.

\noindent\textbf{Dataset Description.} 
All 2D spatial datasets in our evalaution comprise of user check-ins specified as tuples of: user identifier, latitude and longitude of check-in location, and timestamp. Our first public dataset, {\em Gowalla} (GW), is a subset of the user check-ins collected by the SNAP project~\cite{cho2011friendship} from a location based social networking website. It contains 6.4 million records from ~200k unique users during a time period between February 2009 and October 2010. Our second public dataset, SF-CABS-S (CABS)~\cite{piorkowski2009crawdad}, is derived from the GPS coordinates of approximately 250 taxis collected over 30 days in San Francisco. Following \cite{hay2016principled,qardaji2013differentially}, we keep only the start point
of the mobility traces, for a total of 217k records. The proprietary dataset is obtained from Veraset~\cite{verasetref} (VS), a data-as-a-service company that provides anonymized population movement data collected through location measurement signals. It covers approximately 10\% of cellphone devices in the U.S \cite{verasetcoverage}. For a single day in December 2019, there are 2,630,669,304 location signals across the US. Each location signal corresponds to an anonymized device and there are 28,264,106 distinct anonymized devices across the US in that day. Lastly, to further study the impact of different data distributions, we also consider an application-centric processing of the Veraset dataset. We perform Stay Point Detection (SPD) \cite{ye2009mining}, to remove location signals when a person is moving, and to extract POI visits when a user is stationary. SPD is typically useful for POI/location services~\cite{perez2016full}, and results in a data distribution consisting of user visits (meaning fewer points on roads and more at POIs). Following the work of \cite{ye2009mining}, we consider as location visit a region 100 meters wide where a user spends at least 30 minutes.  We call this dataset SPD-VS.

To simulate a realistic urban environment, we focus on check-ins from several cities in the U.S. We group cities into three categories based on their population densities~\cite{pop_density}, measured in people per square mile: \textit{low density} (lower than 1000/sq mi), \textit{medium density} (between 1000 and 4000/sq mi) and \textit{high density} (greater than 4000/sq mi). A total of twelve cities are selected, four in each population density category as listed in Table \ref{tbl:cities}. For each city, we consider a large spatial region covering a $20 \times 20$km$^2$ area centered at [lat, lon]. From each density category we randomly select a \textit{test city} (highlighted in bold in Table \ref{tbl:cities}), while the remaining cities are used as \textit{training cities}. We use the notation <\textit{city}> (<\textit{dataset}>) to refer to the subset of a \textit{dataset} for a particular \textit{city}, e.g., Milwaukee (VS) refers to the subset of VS datasets for the city of Milwaukee.

\noindent\textbf{Experiments on VS. } \noindent\textit{Private dataset}: Our experiments on Veraset can be seen as a case-study of answering RCQs on a proprietary dataset while preserving differential privacy. We evaluate RCQs on the Veraset dataset for the \textit{test cities}. Due to the enormous volume of data, we sample at random sets of $n$ check-ins, for $n\in \{25k, 50k, 100k, 200k, 400k\}$ for the test cities and report the results on these datasets. \textit{Auxiliary Datasets}: For each test city in VS, we set $Q_W$ and $D^*$ to be the GW dataset from the corresponding city. GW and VS datasets are completely disjoint since they are collected almost a decade apart by separate sets of devices. Thus, we can be sure that there is no privacy leakage in the evaluation. Lastly, the public datasets $\mathcal{D}$ are the set of all the training cities of the GW dataset.

\noindent\textbf{Experiments on GW. }\textit{Private dataset}: We present the results on the complete set of records for the test cities of Miami, Milwaukee and Kansas City with 27k, 32k and 54k data points, respectively. \textit{Auxiliary Datasets}: For each test city, we set $Q_W$ and $D^*$ to be the VS counterpart dataset for that city. $\mathcal{D}$ contains all the training cities in the GW dataset. None of the test cities, which are considered sensitive data, are included in $\mathcal{D}$.

\noindent\textbf{Experiments on CABS. } \textit{Private dataset}: Since CABS consists of 217k records within the city of San Francisco only, we treat it as the sensitive test city for publishing. \textit{Auxiliary Datasets}: We set $Q_W$ and $D^*$ to be the GW dataset for San Francisco. $\mathcal{D}$ contains all the training cities in the GW dataset. Once again, collecting auxiliary information from an entirely different dataset ensures no privacy leakage on the considered private dataset.

\begin{table}[t]
\resizebox{\columnwidth}{!}{\begin{tabular}{@{}lll@{}}
\toprule
\multicolumn{1}{c}{Low Pop. density} & \multicolumn{1}{c}{Medium Pop. density} & \multicolumn{1}{c}{High Pop. density} \\ \midrule
Fargo {[}46.877, -96.789{]}              & Phoenix {[}33.448 -112.073{]}               & \textbf{Miami} {[}25.801, -80.256{]}               \\
\textbf{Kansas City} {[}39.09, -94.59{]}          & Los Angeles {[}34.02, -118.29{]}            & Chicago {[}41.880, -87.70{]}             \\
Salt Lake {[}40.73, -111.926{]}          & Houston {[}29.747, -95.365{]}               & SF {[}37.764, -122.43{]}       \\
Tulsa {[}36.153, -95.992{]}              & \textbf{Milwaukee} {[}43.038, -87.910{]}             & Boston {[}42.360 -71.058{]}               \\ \bottomrule
\end{tabular}}
\caption{Urban datasets characteristics.}\label{tbl:cities}
\end{table}


\subsubsection{SNH system parameters}
We first discuss how we set the system parameter $\rho$ using ParamSelect and then present how we set the rest of SNH system parameters. 

\noindent\textbf{ParamSelect for system parameter $\rho$.}
We use the public datasets $\mathcal{D}$ to train the ParamSelect regression model. Note that we train ParamSelect only once, using the set of training cities of the Gowalla Dataset, and utilize the same trained model for all of our evaluated private datasets. In particular, for the nine training cities and five values of privacy budget $\varepsilon$, we obtain 45 training samples. We utilize an AutoML \textit{pipeline}~\cite{yakovlev2020oracle} to choose from among a wide range of ML algorithms. The pipeline splits the training set into J-K (J=3, K=5) cross-validation folds~\cite{moss2018using} to evaluate goodness-of-fit for possible algorithm and hyperparameter combinations. The final selected model is an Extremely Randomized Tree (ExtraTrees)~\cite{geurts2006extremely}, a variant of Random Forest \cite{ho1995random} class of algorithms. ExtraTrees create an ensemble of many random forests, where each tree is trained using the whole learning sample (rather than a bootstrap sample). The final model is trained at a learning rate of 0.1 and its hyperparameters include a total of 150 trees in the forest having a maximum depth of 7. 

\noindent\textbf{Setting other system parameters.}
For other system parameters, we observed that their best value for SNH remains fairly stable over a wide range of different dataset and privacy budget combinations. For example, Secs.~\ref{subsec:hyperparam} and \ref{sec:ablation_uniformity} show this for system parameter $k$ and neural network hyper-parameter, model depth. Simply put, these variables exhibit a simple relationship to final system performance, and when we considered using ParamSelect to set these parameters, we observed little benefit over merely using the values that perform well on any public dataset.

As such, we used one of the cities in the training cities set on Gowalla dataset to decide the system parameter $k$, and the hyper-parameters and architecture of the SNH neural networks. The fully connected neural networks (FCNN) are trained with Adam~\cite{kingma2014adam} optimizer at a learning rate of 0.001. For our application, the performance of a FCNN is stable in the range of [5, 20] for the number of layers and between [40, 100] for width of each layer. Towards the higher range, we noticed an increase in runtime without further gain in accuracy, so we set the FCNN to 20 layers of 80 unit width each. We discuss the impact of model depth on accuracy in Sec.~\ref{subsec:hyperparam}.

\subsubsection{Other experimental settings}\hfill\\
\noindent\textbf{Evaluation Metric.}
We construct query sets of 5,000 RCQs centered at uniformly random positions in the dataset. Each query has side length that varies uniformly from 25 meters to 100 meters. We evaluate the relative error for a query $q$ as defined in Sec.~\ref{sec:prelim}, and set 
smoothing factor $\psi$ to 0.1\% of the dataset cardinality $n$, following previous work \cite{zhang2016privtree,cormode2012differentially,qardaji2013differentially}.


\noindent\textbf{Baselines.}
We evaluate our proposed SNH approach in comparison to state-of-the-art solutions: PrivTree~\cite{zhang2016privtree}, Uniform Grid (UG) \cite{qardaji2013differentially}, Adaptive Grid (AG) \cite{qardaji2013differentially} and Data and Workload Aware Algorithm (DAWA)~\cite{li2014data}. Brief summaries of each method are provided in Sec.~\ref{sec:rel_works}. DAWA requires the input data to be represented over a discrete 1D domain, which can be obtained by applying a Hilbert transformation to 2D spatial data. To obtain the Hilbert curve, we discretize the spatial domain of each dataset into a uniform grid with $2^{20}$ cells, following the work of \cite{li2014data,zhang2016privtree}. Note that DAWA also uses the workload query set, $Q_W$, as specified in Sec.~\ref{exp:setup:dataset}. For PrivTree, we set its fanout to 4, following~\cite{zhang2016privtree}. We also considered Hierarchical methods in 2D (HB2D) \cite{qardaji2013understanding, hay2009boosting} and QuadTree \cite{cormode2012differentially}, but the results were far worse than the above approaches and thus are not reported.

\begin{figure*}
    \centering
\includegraphics[width=\textwidth]{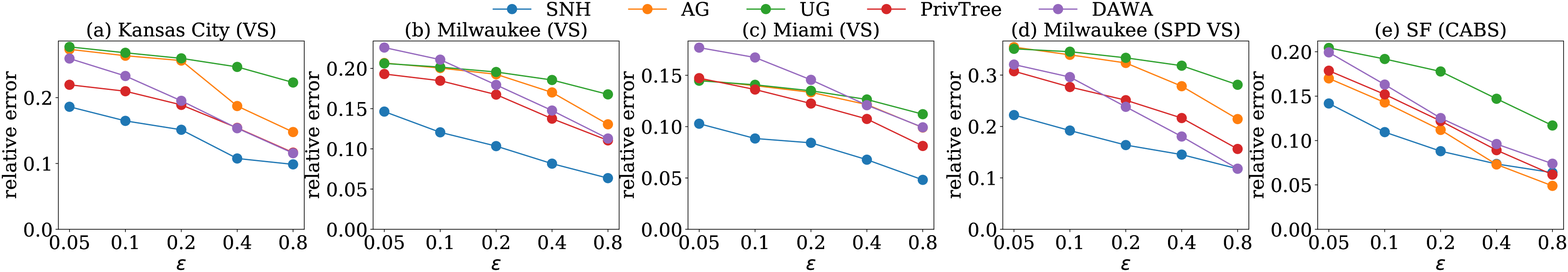}
    \caption{Impact of privacy budget: VS, SPD-VS and CABS datasets}
    \label{fig:vs_cities}
\end{figure*}
\begin{figure*}
 \begin{minipage}{0.59\textwidth}
        \includegraphics[width=\textwidth]{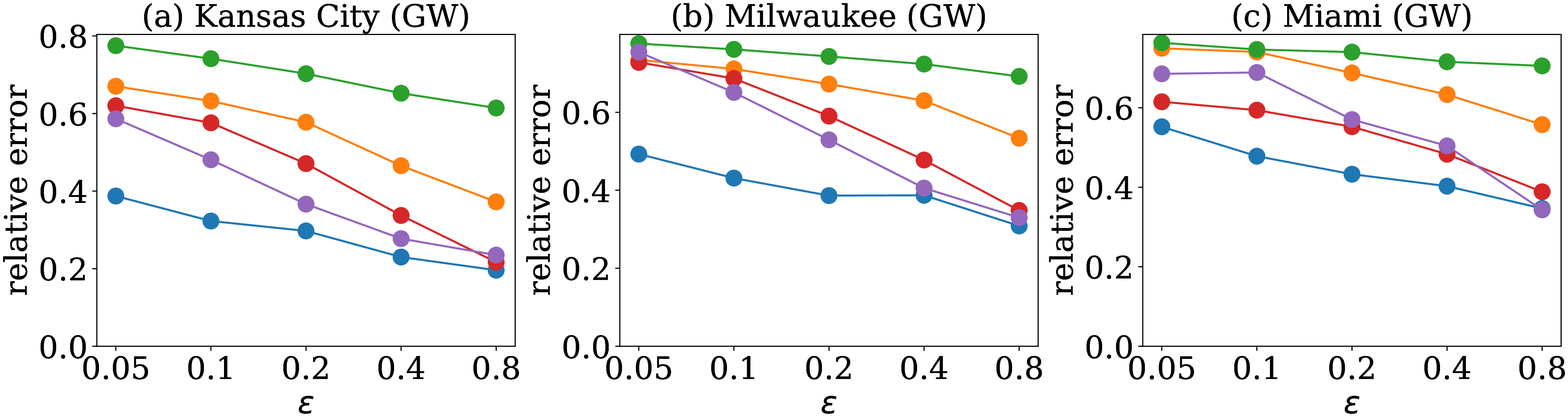}
        \caption{Impact of privacy budget: GW dataset}
        \label{fig:gw_cities}
\end{minipage}
\hfill
\begin{minipage}{0.39\textwidth}
    \includegraphics[width=1\textwidth]{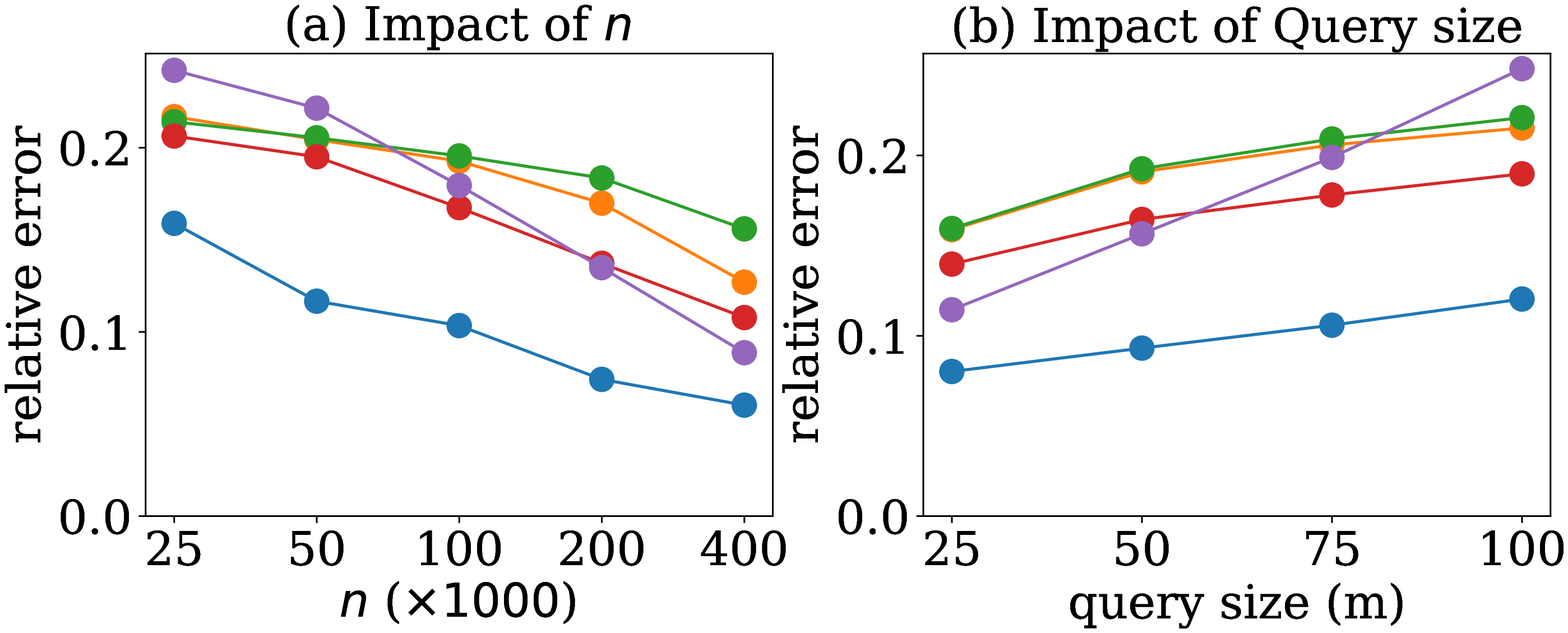}
    \caption{Impact of data and query size}
    \label{fig:vs_data_query}
\end{minipage}
\end{figure*}

\begin{figure*}
    \begin{minipage}[t]{0.59\textwidth}
        \includegraphics[width=\textwidth]{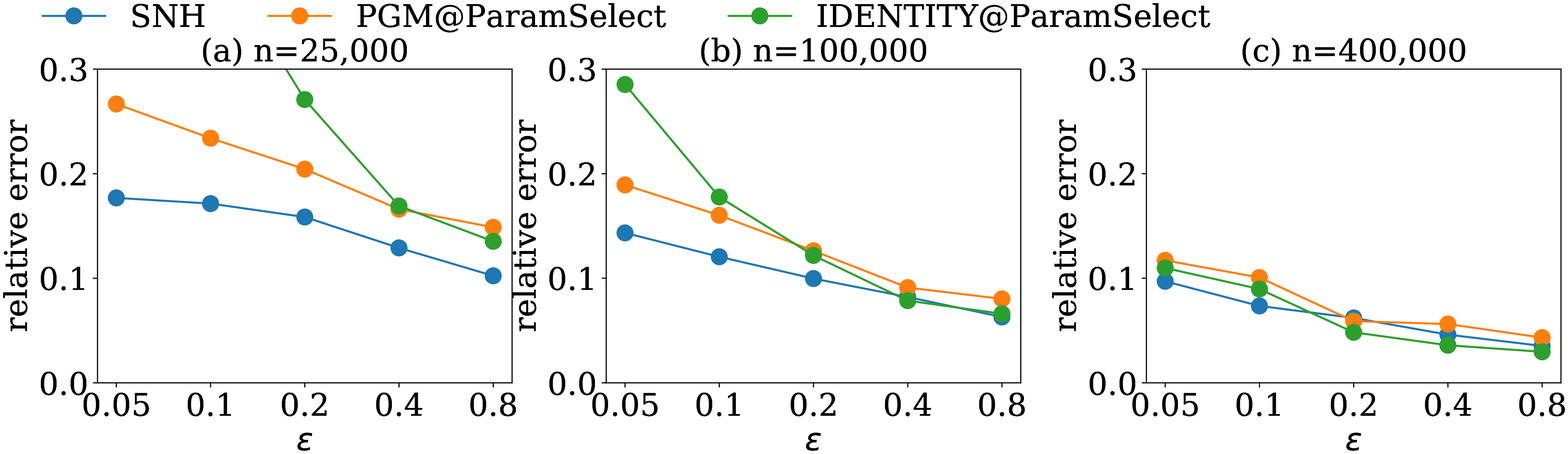}
        \caption{Study of modeling choice}
        \label{fig:modeling}
    \end{minipage}    \begin{minipage}[t]{0.19\textwidth}
        \includegraphics[width=1\textwidth]{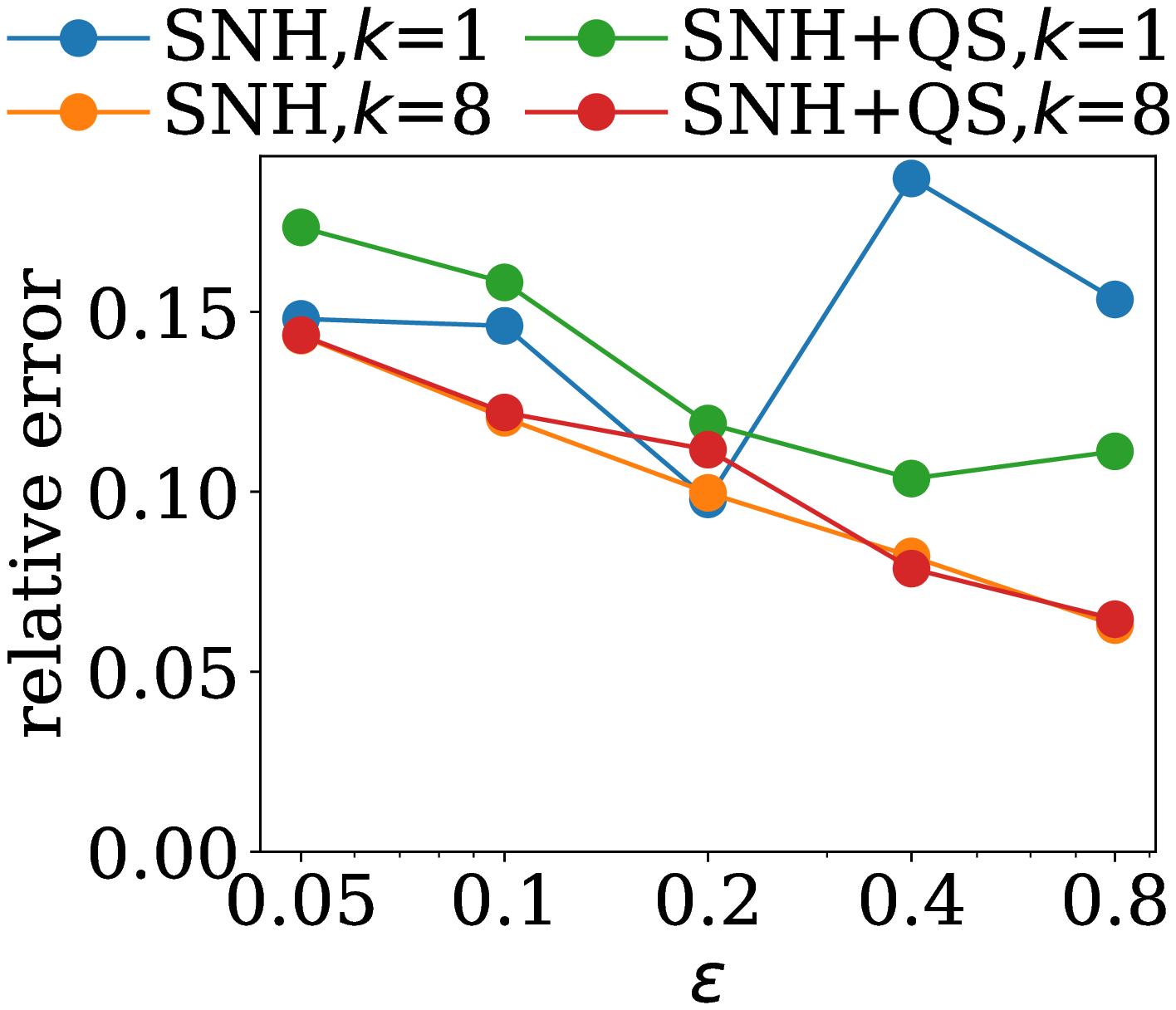}
        \caption{Impact of uniformity assumption}
        \label{fig:uniformity}
    \end{minipage}        \begin{minipage}[t]{0.19\textwidth}
        \includegraphics[width=\textwidth]{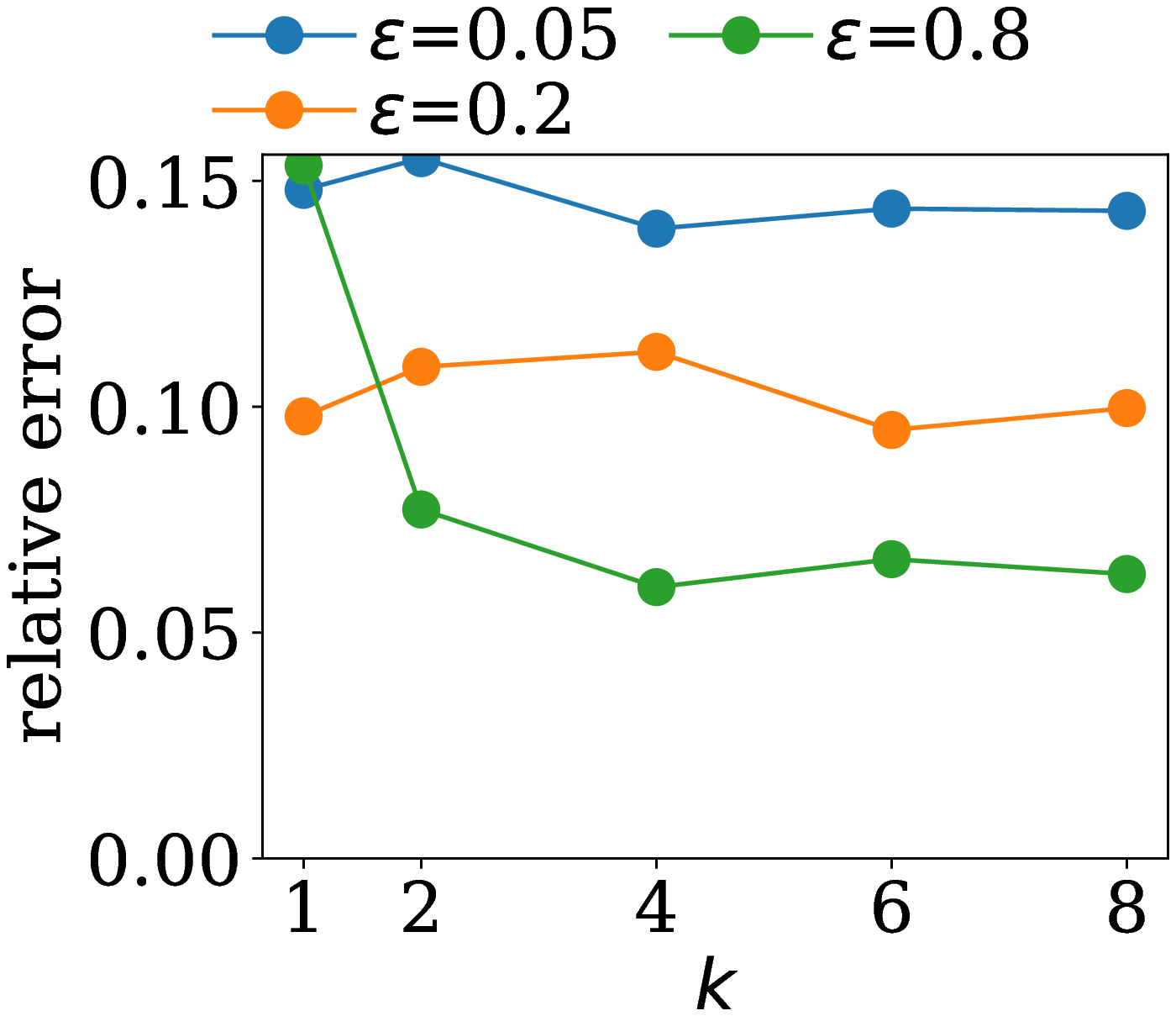}
        \caption{Impact of $k$}
        \label{fig:impact_k}
    \end{minipage}
\end{figure*}

\begin{figure}
\centering

    \begin{minipage}[t]{0.49\columnwidth}
    \vspace{-2cm}
    \begin{tabular}{c|c}
        \textbf{Algorithm} & \textbf{Error (m)}  \\\hline
        ParamSelect & 3.3 \\\hline
        UG & 281.3
    \end{tabular}
    \end{minipage}
        \includegraphics[width=0.19\textwidth]{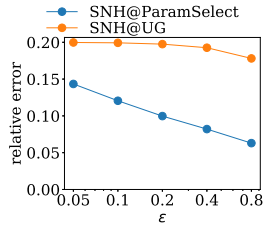}    
    \caption{Study of ParamSelect}
    \label{fig:ablation_paramselect}
\end{figure}
\begin{figure}
    \begin{minipage}{0.4\columnwidth}
    \centering
    \includegraphics[width=\textwidth]{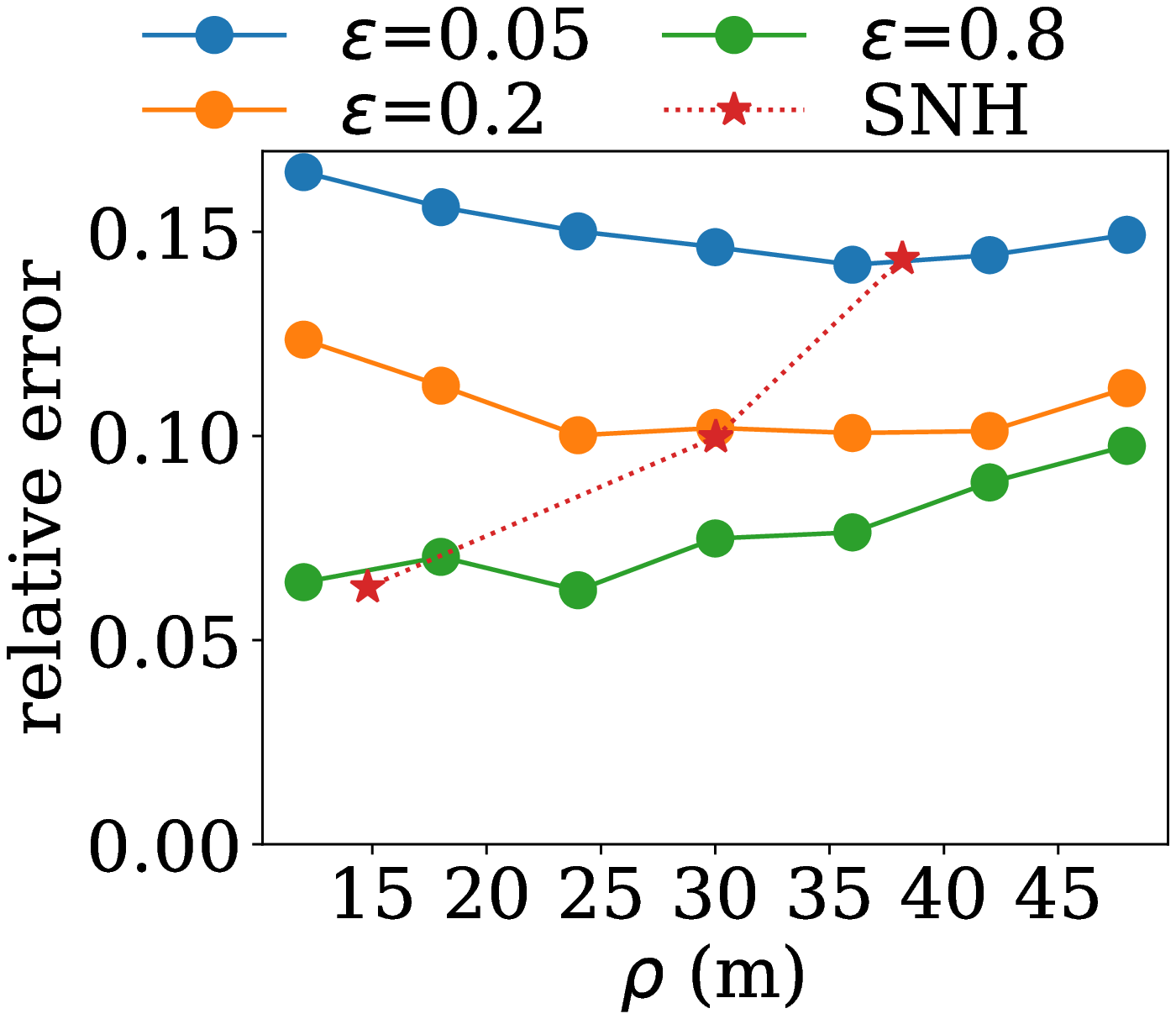}
    \caption{Impact of $\rho$}
    \label{fig:rho}
    \end{minipage}
    \hfill
    \begin{minipage}{0.59\columnwidth}
    \centering
        \includegraphics[width=0.68\textwidth]{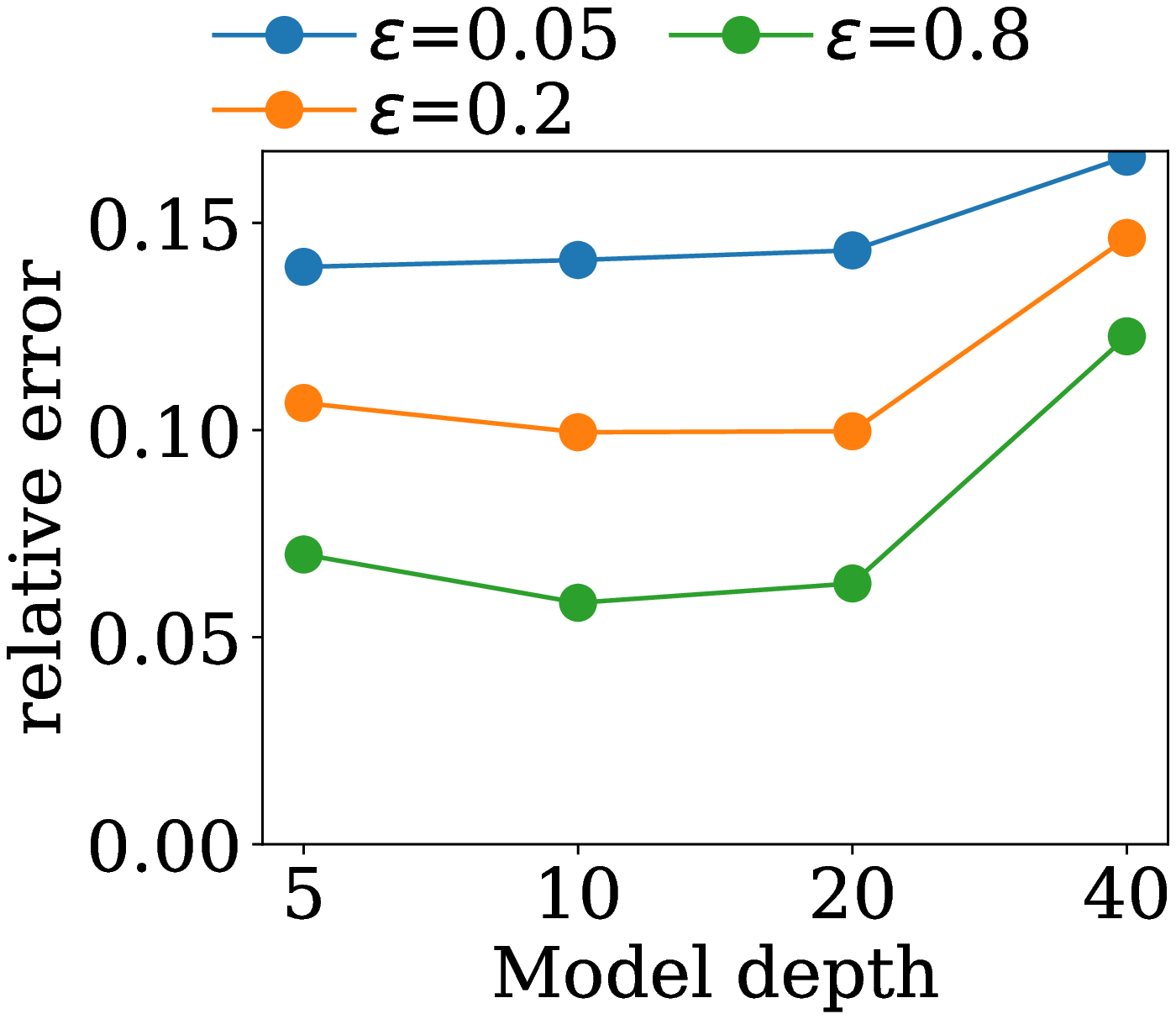}
        \caption{Impact of model depth}
        \label{fig:depth}
    \end{minipage}
\end{figure}
\noindent\textbf{Implementation.}  All algorithms were implemented in Python on Ubuntu Linux 18.04 LTS OS, and executed on an Intel i9-9980XE CPU with 128GB RAM and a RTX 2080 Ti GPU. For all approaches, all data and indices are stored in main memory. Neural networks are implemented in JAX~\cite{jax2018github}, a dedicated language for ML. In the above setup, SNH took up to 20 minutes to train in our experiments, depending on the value of $\rho$. The average query time of SNH is 329$\mu s$ and the model takes 4.06 MB of space. We publicly release the source code at~\cite{snh_implementation}.

\noindent\textbf{Default Values.} Unless otherwise stated, we present the results on the \textit{medium} population density city, Milwaukee (VS), with data cardinality $n= 100k$. Privacy budget $\varepsilon$ is set to $0.2$.

\subsection{Comparison with Baselines}
\label{subsec:comparison}

\noindent\textbf{Impact of privacy budget}. 
Figs.~\ref{fig:vs_cities} and \ref{fig:gw_cities} present the relative error of SNH and competitor approaches at varying degrees of privacy for the test cities in VS, SPD-VS, CABS and GW datasets. Recall that a smaller $\varepsilon$ means stronger privacy protection (hence larger noise magnitude). 

For our proprietary datasets, VS and SPD-VS, we observe that SNH outperforms the state-of-the-art by up to 50\% at all privacy levels (Fig.~\ref{fig:vs_cities} (a)-(d)). This shows that SNH is effective in utilizing machine learning and publicly available data to improve accuracy of privately releasing proprietary datasets. Fig.~\ref{fig:vs_cities} (e) and Fig.~\ref{fig:gw_cities}  show that SNH also outperforms the benchmarks for CABS and GW datasets in almost all settings, the advantage of SNH being more pronounced at stricter privacy regimes (i.e., at smaller $\varepsilon$). We note that stricter privacy regimes are particularly important for location data, since such datasets are often released at multiple time instances, hence the privacy budget per release gets smaller. Overall, our results show that SNH is robust to different types of location data, with the ability to perform well on cell phone location signals (in VS), user staypoints (in SPD-VS), taxi locations (in CABS) and user check-ins (in GW) datasets across different cities.



 
\noindent\textbf{Impact of data cardinality}. Fig.~\ref{fig:vs_data_query} (a) shows the impact of data cardinality on relative error for Milwaukee (VS). For all algorithms, the accuracy improves as data cardinality increases. This is a direct consequence of the signal-to-noise ratio improving as cell counts are less impacted by DP noise. SNH consistently outperforms competitor approaches at a wide range of data cardinality settings.

\noindent\textbf{Impact of query size}.
We evaluate the impact of query size on accuracy by considering test queries of four different sizes in Milwaukee (VS). Fig.~\ref{fig:vs_data_query} (b) shows that the error for all the algorithms increases when query size grows, with SNH outperforming the baselines at all sizes. There are two competing effects when increasing query size: on the one hand, each query is less affected by noise, since actual counts are larger; on the other hand, the error from more grid cells is aggregated in a single answer. The second effect is stronger, so the overall error steadily increases with query size.\\

Overall, results on both public and proprietary data show SNH outperforming competitor techniques for different cities, data sizes and query sizes. In the remainder of this section, we no longer consider benchmarks, and focus on studying the behavior of SNH when varying its system parameters.




\subsection{Ablation study}\label{subsec:ablation}
\subsubsection{Modeling choices} Recall that SNH first creates a uniform grid, with granularity decided by ParamSelect. It then performs data augmentation and learning using the data collected on top of the grid. Next, we study the importance of each component of SNH to its overall performance. We create two new baselines to show how our choice of using neural networks to learn the patterns in the data improves performance. 
The first, called IDENTITY@ParamSelect, ablates SNH, utilizing only the uniform grid created in SNH at data collection.
The second baseline, called PGM@ParamSelect, employs Private Probabilistic Graph Models (PGM) \cite{mckenna2019graphical}, a learning algorithm specifically designed for high-dimensional categorical data. We extend PGM to 2D spatial datasets by feeding it a DP uniform grid at the granularity selected by ParamSelect.

Fig.~\ref{fig:modeling}(a) shows SNH outperforming both these baselines. However, the benefit diminishes when the privacy budget and the data cardinality increase (note that both $n$ and $\varepsilon$ are in log scale), where a simple uniform grid chosen at the \textit{correct} granularity outperforms all existing methods (comparing Fig.~\ref{fig:modeling}(b) with Fig.~\ref{fig:vs_cities}(b) shows IDENTITY@ParamSelect outperforms the state-of-the-art for $\varepsilon=0.4$ and 0.8). 
For such ranges of privacy budget and data cardinality, ParamSelect recommends a very fine grid granularity. Thus, the uniformity error incurred by IDENTITY@ParamSelect becomes lower than that introduced by the modelling choices of SNH and PGM. This also shows the importance of a good granularity selection algorithm, as UG in Fig.~\ref{fig:vs_cities} performs worse than IDENTITY@ParamSelect for larger values of $\varepsilon$ (recall that both UG and IDENTITY@ParamSelect are uniform grids, but with different granularities).

\subsubsection{Balancing Uniformity Errors}\label{sec:ablation_uniformity} We discuss how the use of the uniformity assumption at different stages of SNH impacts its performance. Recall that the uniformity assumption is used in data augmentation (pre-learning) and model utilization (post-learning) steps, with the value of $k$ balancing the use of the uniformity assumption pre- and post-learning. That is, when $k$ is larger, the uniformity assumption is used more often pre-learning, to create more training data. When $k$ is smaller, the uniformity assumption is used more prominently post-learning as larger parts of the space are assumed to be uniformly distributed to answer a query (and thus the ratio of $\frac{r}{r^*}$ is further away from $1$).

Furthermore, we study how removing the uniformity assumption post-learning, and replacing it with a neural network, affects the performance. Specifically, we consider a variant of SNH where we train the neural networks to also take as an input the query size. Each neural network is still responsible for a particular set of query sizes, $[r_l, r_u]$, where we use data augmentation to create query samples with different query sizes falling in $[r_l, r_u]$. Subsequently, instead of scaling the output of the trained neural networks, now each neural network also takes the query size as an input, and thus, the answer to a query is just the forward pass of the neural network. We call this variant \textit{SNH with query size}, or SNH+QS .

Figs.~\ref{fig:uniformity} and \ref{fig:impact_k} show the results of this set of experiments. Fig.~\ref{fig:uniformity} shows that, first, removing the uniformity assumption post-learning has almost no impact on accuracy when $k$ is large. However, for a small value of $k$, it provides more stable accuracy. Note that when $k=1$, SNH trains only one neural network for query size $r^*$ and answers the queries of size $r$ by scaling the output of the neural network by $\frac{r}{r^*}$. The error is expected to be lower when $\rho$ and $r^*$ have similar values, since there will be less uniformity error when performing data augmentation. This aspect is captured in Fig.~\ref{fig:uniformity}, where at $\varepsilon=0.2$, $r^*$ and $\rho$ are almost the same values and thus, the error is the lowest.  

Finally, Fig.~\ref{fig:impact_k} shows the impact of $k$ on the accuracy of the models. The result shows that for large values of $\varepsilon$, increasing $k$ can substantially improve the performance. 

\subsubsection{Benefit of ParamSelect}\label{sec:exp:paramselect} ParamSelect finds the best grid granularity by learning from a public dataset. An existing alternative for setting the grid granularity is using the guideline of UG \cite{qardaji2013differentially}, which, by making assumptions about the query and data distribution, analytically formulates the error for using a uniform grid. It then proposes creating an $m\times m$ grid, setting $m=\sqrt{n\varepsilon/c}$ for a constant $c$ empirically set to $c=10$. We call SNH with grid granularity chosen this way SNH@UG. We compare this method with SNH (referred to SNH@ParamSelect to emphasize that ParamSelect was used to set the value of $\rho$).

Fig.~\ref{fig:ablation_paramselect} shows the result of this experiment. First, the table in Fig.~\ref{fig:ablation_paramselect} (left) shows the mean absolute error of the predicted cell width for the grid. This is measured by first empirically finding $\rho^*$, the cell width at which SNH achieves highest accuracy. The absolute error when either UG or ParamSelect suggest cell width $\rho$, is then $|\rho-\rho^*|$. The error in Fig.~\ref{fig:ablation_paramselect} (left) is the average of this error for different privacy budgets.  

Second, Fig.~\ref{fig:ablation_paramselect} (right) shows how different predicted cell width values impact the accuracy of SNH. We observe a significant difference between SNH@UG and SNH@ParamSelect, establishing the benefits of ParamSelect. Overall, the results of this ablation study, and the ablation study in Sec.~\ref{sec:ablation_uniformity}, show that both good modelling choices and system parameter selection are imperative in order to achieve high accuracy.



\subsection{System parameters
analysis}\label{subsec:hyperparam}

\noindent\textbf{Impact of $\rho$}. Fig.\ref{fig:rho} shows the performance of SNH with varying cell width $\rho$, at multiple values of $\varepsilon$. In all instances, a U-shaped trend is observed. Increasing the grid granularity at first improves accuracy by improving the average signal-to-noise ratio at each cell, however a grid too coarse reduces the number of training samples that can be extracted for training SNH to answer queries of various sizes. Given a dataset cardinality, this trade-off shifts to smaller values of $\rho$ for larger values of $\varepsilon$ as the lower DP noise impacts less aggressively cell counts. Lastly, the red line in Fig.\ref{fig:rho} labelled SNH, shows the result of SNH at the granularity chosen by ParamSelect. SNH performing close to the best possible proves that ParamSelect finds an advantageous cell width for SNH. 

\noindent\textbf{Impact of Model Depth}. We study how the neural network architecture impacts SNH's performance in Fig.\ref{fig:depth}. Specifically, we vary the depth (i.e., the number of layers) of the network. Increasing model depth improves slightly the accuracy of SNH due to having better expressive power from deep neural networks. However, networks that are too deep quickly decrease accuracy as the gradients during model training diminish dramatically as they are propagated backward through the very deep network. Furthermore, larger $\varepsilon$ values are able to benefit more from the increase in depth, as more complex patterns can be captured in the data when it is less noisy.

\section{Related Work}\label{sec:rel_works}
\noindent\textbf{Privacy preserving machine learning}. A learned model can leak information about the data it was trained on \cite{shokri2017membership,hitaj2017deep}. 
Recent efforts focused on training differentially private versions of machine learning algorithms, including empirical risk minimization (ERM) \cite{chaudhuri2011differentially, kifer2012private} and deep neural networks \cite{sealfon2021efficiently, abadi2016deep}. Based on how noise is injected, such approaches can be categorized as output perturbation \cite{wu2017bolt} (i.e., after training), objective perturbation~\cite{chaudhuri2011differentially, kifer2012private}, which adds to the objective function a random regularization term, 
and gradient perturbation algorithms~\cite{abadi2016deep}, which add noise to the gradient of the loss function during training. 
The first two approaches ensure privacy for algorithms that optimize a convex objective such as ERM, while gradient perturbation guarantees differential privacy  even for non-convex objectives, e.g., in deep neural networks. Our approach is different in that we sanitize the training data {\em before} learning. Furthermore, the work of \cite{abadi2016deep} achieves ($\varepsilon, \delta$)-DP~\cite{nissim2007smooth, erlingsson2014rappor, censusgovKDD18}, a weaker guarantee than $\varepsilon$-DP which is achieved by our method.

\noindent\textbf{Answering RCQs}. The task of answering RCQs is well explored in the literature. Classes of algorithms can be categorized based on data dimensionality and their data-dependent or independent nature. However there is no single dominant algorithm for all domains\cite{hay2016principled}. 

In the one dimensional case, the data-independent Hierarchical method \cite{hay2009boosting} uses a strategy consisting of hierarchically structured range queries typically arranged as a tree. Similar methods (e.g., HB  ~\cite{qardaji2013understanding}) differ in their approach to determining the tree's branching factor and allocating appropriate budget to each of its levels. Data-dependent techniques, on the other hand, exploit the redundancy in real-world datasets in order to boost the accuracy of histograms. The main idea is to first lossly compress the data. For example, EFPA \cite{acs2012differentially} applies the Discrete Fourier Transform  whereas DAWA~\cite{li2014data} uses dynamic  programming to compute the least cost partitioning. The compressed data is then sanitized, for example, directly with Laplace noise~\cite{acs2012differentially} or with a greedy algorithm that tunes the privacy budget to an expected workload~\cite{li2014data}).  

While some approaches such as DAWA and HB extend to 2D naturally, others specialize to answer spatial range queries. Uniform Grid (UG) \cite{qardaji2013differentially} partitions the domain into a $m \times m$ grid and releases a noisy count for each cell. The value of $m$ is chosen in a data-dependent way, based on dataset cardinality. Adaptive Grid (AG) \cite{qardaji2013differentially} builds a two-level hierarchy: the top-level partitioning utilizes a granularity coarser than UG. For each bucket of the top-level partition, a second partition is chosen in a data-adaptive way, using a finer granularity for regions with a larger count. QuadTree~\cite{cormode2012differentially} first generates a quadtree, and then employs the Laplace mechanism to inject noise into the point count of each node. Range-count queries are answered via a top-down traversal of the tree. Privtree \cite{zhang2016privtree} is another hierarchical method that allows variable node depth in the indexing tree (as opposed to fixed tree heights in AG, QuadTree and HB). It utilizes the Sparse-Vector Technique \cite{svtvldb2017} to determine a cell's density prior to splitting the node.


The case of high-dimensional data was addressed by~\cite{mckenna2018optimizing, xiao2012dpcube, zhang2017privbayes}. The most accurate algorithm in this class is High-Dimensional Matrix Mechanism (HDMM) \cite{mckenna2018optimizing} which represents queries and data as vectors, and uses sophisticated optimization and inference techniques to answer RCQs. An inference step solves an unconstrained ordinary least squares problem to minimize the $L_2$ error on the input workload of linear queries. 
PrivBayes~\cite{zhang2017privbayes} is a mechanism that privately learns a Bayesian network over the data in order to generate a synthetic dataset that can consistently answer workload queries. Budget allocation is equally split over learning the Bayesian network structure and learning its network parameters. Due to the use of sampling to estimate data distribution, it is a poor fit for skewed spatial datasets. Most similar to our work is PGM \cite{mckenna2019graphical}, which utilizes Probabilistic Graphical Models to measure a compact representation of the data distribution, while minimizing a loss function. Data projections over user-specified subgroups of attributes are sanitized and used to learn the model parameters. PGM is best used in the inference stage of modelling sensitive high-dimensional datasets with many categorical attributes. It improves accuracy of privacy mechanisms (such as HDMM and PrivBayes) that can already capture a good model of the data.


\noindent\textbf{Private parameter tuning}
The process of determining the system parameters of a private data representation must also be DP-compliant. Several approaches utilize the data themselves to tune system parameters such as depth of a hierarchical structure (e.g., in QuadTree or HB) or spatial partition size (e.g. k-d trees), without consideration of privacy~\cite{hay2009boosting}. In particular, UG~\cite{qardaji2013differentially} models the grid granularity as $m=\sqrt{n\varepsilon/c}$, where the authors tune $c=10$ empirically on 
sensitive datasets, in breach of differential privacy. Moreover, this results in the value of $c$ overfitting their structures to test data, and result in poor utility on new datasets~\cite{hay2009boosting}. The generalization ability of system parameters is also poor when parameter selection is performed arbitrarily, independent of data (as show by Hay et. al.~\cite{hay2009boosting} in their DP-compliant adoption of HB and QuadTree). Using \textit{public} datasets to tune system parameters of domain partitioning or learning models is a better strategy~\cite{chaudhuri2011differentially}. We note that our attempt to determine an advantageous cell width for a differentially-private grid is similar in effort to that in UG~\cite{qardaji2013differentially}. However, our proposed hyper-parameter strategy of parameter selection vastly improves generalization ability over UG~\cite{qardaji2013differentially} by exploiting additional dataset features and their non-linear relationships. 




\section{Conclusion}\label{sec:conclusion}
We  proposed SNH: a novel method for answering range count queries on location datasets while preserving differential privacy. To address the shortcomings of existing methods (i.e., over-reliance on the uniformity assumption and noisy local information when answering queries), SNH utilizes the power of neural networks to learn patterns from location datasets. We proposed a two stage learning process: first, noisy training data is collected from the database while preserving differential privacy; second, models are trained using this sanitized dataset, after a data augmentation step. In addition, we devised effective machine learning strategies for tuning system parameters using only {\em public} data. Our results show SNH outperforms the state-of-the-art on a broad set of input  data with diverse characteristics. In future work, we plan to extend our solution to the releasing of user trajectories datasets, which are more challenging due to the higher data dimensionality and increased sensitivity of sequential data.

\bibliographystyle{ACM-Reference-Format}
\bibliography{references} 

\end{document}